\documentclass[twocolumn,article,superscriptaddress,showpacs]{revtex4-1}

\usepackage[utf8]{inputenc}
\usepackage{color}
\usepackage{amsmath}
\usepackage{amssymb}
\usepackage{graphicx}
\usepackage{esint}
\usepackage[unicode=true,
 bookmarks=false,
 breaklinks=true,pdfborder={0 0 0},backref=false,colorlinks=true]
 {hyperref}

\makeatletter

\usepackage{epstopdf}\usepackage{subfigure}
\usepackage{float}
\usepackage{amsfonts}
\usepackage{bm}% bold math
\usepackage{subfigure}
\usepackage{braket}

\def\MSA{MoSi$_2$As$_4$}
\def\MS{MoS$_2$}

\makeatletter

\usepackage{amsfonts}
\usepackage{bm}% bold math
\usepackage{braket}

\makeatother

\begin{document}

\title{{Chiral phonons entangled with multiple Hall effects and unified convention for pseudoangular momentum in 2D materials}
}

\author{Tiantian Zhang}
\affiliation{Department of Physics, Tokyo Institute of Technology, Ookayama, Meguro-ku, Tokyo 152-8551, Japan}
\affiliation{Tokodai Institute for Element Strategy, Tokyo Institute of Technology, Nagatsuta, Midori-ku, Yokohama, Kanagawa 226-8503, Japan}
\author{Shuichi Murakami}
\affiliation{Department of Physics, Tokyo Institute of Technology, Ookayama, Meguro-ku, Tokyo 152-8551, Japan}
\affiliation{Tokodai Institute for Element Strategy, Tokyo Institute of Technology, Nagatsuta, Midori-ku, Yokohama, Kanagawa 226-8503, Japan}

\begin{abstract}
Recently, a series of two-dimensional~(2D) nonmagnetic layered materials $X$Si$_2Y_4$ ($X$=transition metals; $Y$=pnictogens) having similar crystal structures with transition-metal dichalcogenides~(TMDs) were proposed for their potential application value. 
Like TMDs, we propose that chiral phonon involved valley-selective optical circular dichroism can be also obtained in $X$Si$_2Y_4$, and it can be further entangled with multiple Hall effects. 
However, it is difficult to compare such effect between $X$Si$_2Y_4$ and TMDs due to the non-unified conventions for pseudo-angular momentum (PAM) in the previous studies. 
Here we use \MSA{} and \MS{} as examples to establish unified convention for both phonon PAM and electronic PAM, together with showing their similarities and differences in crystal structure, band structures and valley-selective optical circular dichroism. 
In particular, we find in \MSA{}, the chiral phonon emission/absorption, the spin, the valley and the transverse Hall current, can be modulated by changing either the energy or the handedness of the incident light, which show better performance than TMDs. 

%(150words)
\end{abstract}

\maketitle

%---------------------------------------------introduction-----------------------
\section*{Introduction}
Two-dimensional~(2D) materials have attracted increasing interests due to their intriguing properties and versatile applications that emerge in the monolayer limit. For example, TMDs were proposed as promising 2D materials for valleytronic and spintronic applications years ago, but the limited size and low carrier mobility are obstacles for their further applications~\cite{feldman1995high,wang2015all,chang2020fast,radisavljevic2011single}. 
Recently, a series of 2D nonmagnetic van der Waals (vdW) layered materials $X$Si$_2Y_4$ ($X$=transition metals; $Y$=nitrogen group elements) having the same crystal symmetries with TMDs were proposed as alternatives, benefiting from their excellent properties like synthesized with large size, good ambient stability, considerable hole mobility, and a wide range band gaps within visible light range~\cite{Hong2020,Novoselov2020,li2020valley,chen2021first,Yao2021,yang2021accurate,wu2021semiconductor,zhong2021strain,Bafekry_2021,kang2021second,cui2021tuning}. They also show intrinsic piezoelectricity~\cite{guo2020intrinsic,mortazavi2021exceptional,GUO2021110223,LI2021114753}, strongly suppressed Fermi level pinning, wide-range tunable Schottky barrier height~\cite{wang2021article,cao2021two}, high intrinsic lattice thermal conductivity~\cite{Yu2021,mortazavi2021exceptional} and topological property~\cite{wang2021intercalated} by first-principle calculations, which make them practical platforms for 2D electronic and thermal devices~\cite{nandan2021two}. 
In addition, due to their multivalley electronic band structure, physical processes induced by the valley degree of freedom can be also realized in $X$Si$_2Y_4$, such as valley polarization \cite{zeng2012valley,carvalho2017intervalley,cao2012valley,wang2012electronics,he2014tightly,wu2013electrical,jones2013optical,xu2014spin} and photon-involved intervalley scattering~\cite{chen2021probing,drapcho2017apparent,tatsumi2018interplay,malard2009raman,zhu2018observation,chen2019entanglement,li2019momentum,li2019emerging}.

{Phonons in $X$Si$_2Y_4$ and TMDs also have multivalley band structures, and they have pseudo-angular momentum (PAM). The PAM is a quantum number characterizing symmetry properties of phonons, and it is different from the angular momentum. 
Material samples are always kept in still in experiments, which prevents the angular momentum from conserving within the sample, whereas PAM can still be conserved. Such conservation for PAM will bring valley-selective circular dichroism in materials like TMDs, associated with chiral phonons. We note that the definition of PAM is analogous to that of pseudomomentum, and their similarities and differences are discussed in detail in Ref.~\cite{simon2021difference}.} 
Although studies on chiral phonons have been widely implemented in TMDs via circular dichroism of interband transitions, most of them are under hole doping conditions~\cite{zhang2015chiral,chen2021propagating,zhang2021chiral}. 
{In addition, there is also no unified convention proposed for both electron PAM ($l_{e}$) and phonon PAM ($l_{ph}$); in particular the definitions for their Bloch-phase ($l^o$) and self-rotation ($l^s$) part of PAM were confusing in different works.} 
Thus, even for the same physical process, it is difficult to compare between different materials within TMD family, not to mention to compare with the materials beyond TMDs.

%Definitions on PAM, they will not only make people hard to compare the same physical process between different materials, $e.g.$, intervalley scattering effects in 2D TMDs and transition-metal compounds with similar structures, but also make the readers confused for the inconsistent results caused by different convention. 

In this paper, by comparing the typical TMD material \MS{} with one of the $X$Si$_2Y_4$ material \MSA{}, we establish the unified convention for the phonon PAM and electron PAM in the chiral phonon involved intervalley scattering process. {We also systematically investigate their similarities and differences in crystal and electronic band structures. }
{\MS{} is an indirect-gap semiconductor in the multilayer limit, while it will become a direct band-gap one at monolayer~\cite{wang2012electronics,splendiani2010emerging}. 
\MSA{} has a direct band gap, which is irrespective of the layer number.}  
By unifying the convention with \MSA{} and \MS{}, we can fix the confusing results in previous papers for WN$_2$~\cite{chen2021propagating} and obtain a unified picture for all materials having similar crystal structures with TMDs. 
Furthermore, like TMDs, \MSA{} only have $C_3$-invariant Wyckoff positions (WPs) in its crystal structure. 
It means that phonon PAM and phonon angular momentum are equal for the phonon modes at $\Gamma$, and that in \MSA{}, both the phonon PAM and chiral phonon vibration in the real space at the $K$ and $K^{\prime}$ valleys are strongly constrained by symmetry~\cite{zhang2021chiral}, like all the other materials in 2D TMDs.

%%
%---------------------------------------------Fig. 1-----------------------
\begin{center}
\begin{figure}[!tp]
\includegraphics[scale=0.6]{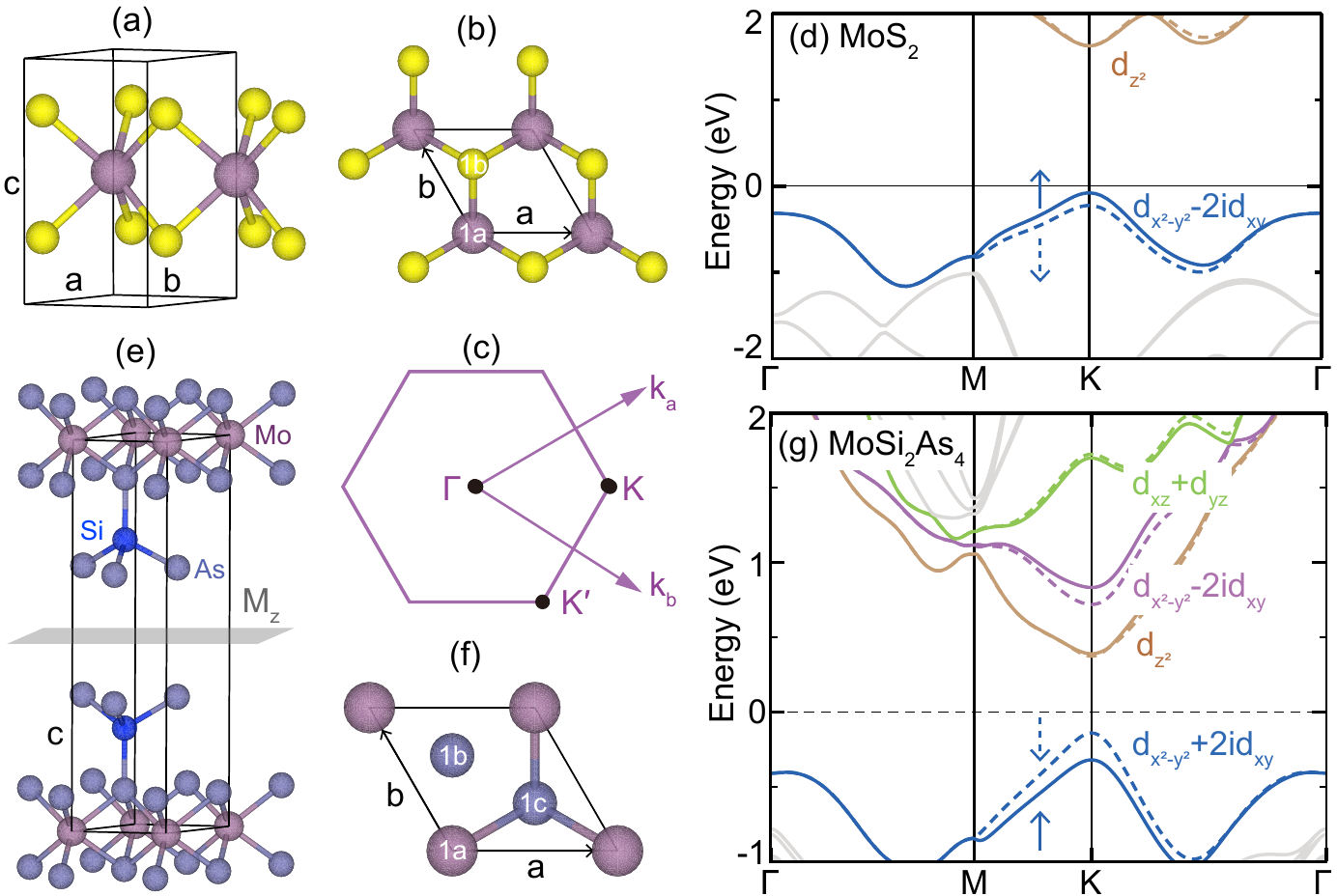}\caption{(a-b) Side view and top view of the crystal structure for \MS, which only have two $C_3$-invariant Wyckoff positions in the 2D limit, i.e., $1a$ and $1b$. (c) is the Brillouin zone for both \MSA{} and \MS{}. 
(e-f) Side view and top view of the crystal structure for \MSA{}, which has three $C_3$-invariant Wyckoff positions in the 2D limit, i.e., $1a$, $1b$ and $1c$.   (d) and (g) are the spinful electronic band structure for \MS{} and \MSA, respectively. Dashed lines represent the spin-down states, while the solid lines represent the spin-up states. 
\label{fig:aa}}
\end{figure}
\end{center}
%----------

\begin{table*}[]
\begin{tabular}{c|cc|cccc|cccc|cccc}
\hline
\multicolumn{1}{c|}{}   & \multicolumn{2}{c|}{prism} & \multicolumn{4}{c|}{VBM$-1$}                                                                                                                                                                                                                             & \multicolumn{4}{c|}{VBM}                                                                                                                                                                                                                             & \multicolumn{4}{c}{CBM}                                                                                                                                                                                                                              \\ \hline
item                    & WP            & WP         & \begin{tabular}[c]{@{}c@{}}orbital \\ at $K$\end{tabular} & \multicolumn{1}{c|}{\begin{tabular}[c]{@{}c@{}}orbital \\ at $K^{\prime}$\end{tabular}} & \begin{tabular}[c]{@{}c@{}}spin \\ at $K$\end{tabular} & \begin{tabular}[c]{@{}c@{}}spin \\at $K^{\prime}$\end{tabular} & \begin{tabular}[c]{@{}c@{}}orbital \\ at $K$\end{tabular} & \multicolumn{1}{c|}{\begin{tabular}[c]{@{}c@{}}orbital \\ at $K^{\prime}$\end{tabular}} & \begin{tabular}[c]{@{}c@{}}spin\\ at $K$\end{tabular} & \begin{tabular}[c]{@{}c@{}}spin \\at $K^{\prime}$\end{tabular} & \begin{tabular}[c]{@{}c@{}}orbital \\ at $K$\end{tabular} & \multicolumn{1}{c|}{\begin{tabular}[c]{@{}c@{}}orbital \\ at $K^{\prime}$\end{tabular}} & \begin{tabular}[c]{@{}c@{}}spin\\at $K$ \end{tabular} & \begin{tabular}[c]{@{}c@{}}spin\\at $K^{\prime}$\end{tabular} \\ \hline
\MS{}  & $1a$(Mo)           &  $1b$(S)          & $D_{xy}^{-}$                                                 & \multicolumn{1}{c|}{$D_{xy}^{+}$}                                                   & up                                                   & down                                                    & $D_{xy}^{-}$                                                  & \multicolumn{1}{c|}{$D_{xy}^{+}$}                                                   & down                                                  & up                                                   & $d_{z^2}$                                                     & \multicolumn{1}{c|}{$d_{z^2}$}                                                      & up                                                  & down                                                   \\
\MSA{} & $1a$(Mo)           & $1c$(As)          & $D_{xy}^{+}$                                                  & \multicolumn{1}{c|}{$D_{xy}^{-}$ }                                                   & down                                                   & up                                                    & $D_{xy}^{+}$                                                   & \multicolumn{1}{c|}{$D_{xy}^{-}$ }                                                   & up                                                  & down                                                   & $d_{z^2}$                                                     & \multicolumn{1}{c|}{$d_{z^2}$}                                                      & down                                                  & up                                                   \\ \hline
\end{tabular}
\caption{Wyckoff positions and valley polarizations for \MS{} and \MSA{}, including the Wyckoff positions for the atoms composing the prism, $l_e^o$, $l_e^s$ and $l_e$ for the three bands near the Fermi level at both valley $K$ and valley $K^{\prime}$. ``$D_{xy}^{\pm}$'' represent the orbitals of $d_{x^2-y^2}\pm2id_{xy}$. ``VBM$-1$'', ``VBM'' and ``CBM'' are the second valence band maximum, valence band maximum and conduction band minimum, respectively. 
Due to the $\frac{\pi}{3}$ rotation on the prism between \MSA{} and \MS{},  
the valley $K$ ($K^{\prime}$) in \MS{} is equal to the valley $K^{\prime}$ ($K$) in \MSA{}. Considering this correspondence of the valleys, we see that the properties of valleys are identical between the two materials.  
}
\label{table:cmp}
\end{table*}

%However, the Bloch-phase part and self-rotation part of PAM in the previous studies do not have a unified convention, especially the ones for electrons are confusing, which will be not easy to compare the physical phenomena for materials with similar crystal structures, such as TMDs and other transition-metal compounds. 
%Thus, we will use \MSA{} as an example to unify the definition and convention for PAM, including both phonons and electrons, by comparing with \MS{}. Our work lay a convention foundation for the related work in the future, and offer an ideal candidate for applications. 

\section{Comparison between \MSA{} and \MS{}}
For most of the 2D materials, their monolayer structures are obtained by exfoliation of their corresponding 3D crystals. 
However, \MSA{} is a 2D material without 3D layered parent structure, which provides huge opportunity for its functional application. 
\MSA{} and TMDs have similar hexagonal structures and physical phenomena. 
In the following, we will show the similarities and differences between \MSA{} and TMDs (with \MS{} as an example), in particular on the crystal structure, electronic bands (Table~\ref{table:cmp}) and chiral phonon involved intervalley scattering process (Table~\ref{table:pam}).

\subsection{Crystal band structures for \MSA{} and \MS}
\MSA{} and \MS{} have the same space group of \#187, and both of them only have $C_3$-invariant Wyckoff positions (WPs) in their crystal structures.  
As shown in Figs.~\ref{fig:aa} (a-b), \MS{} only has a prism (\MS) crystal structure with Mo atom located at WP $1a$ and S atom located at $1b$. We note that all the 2D TMDs and hexagonal binary compounds share the same crystal structure with \MS{}, which will lead to similar physical phenomena in those materials. 
As for \MSA{} shown in Figs.~\ref{fig:aa} (e-f), the monolayer is built up by septuple atomic layers of As (WP $1b$)-Si (WP $1c$)-As (WP $1c$)-Mo (WP $1a$)-As (WP $1c$)-Si (WP $1c$)-As (WP $1b$), which can be treated as a MoAs$_2$ prism layer sandwiched between two triangular pyramid layers consisting of As (WP $1b$) and Si (WP $1c$). 
Those two triangular pyramids are related by $M_z$ symmetry. 

%%%-------- 
Crystal structures of those two materials are similar, with correspondence between the S atoms in \MS{} and the As atoms around Mo atoms in \MSA{}. 
{Meanwhile, the unit-cell conventions of these two materials widely used in literatures are different} and are related by $\frac{\pi}{3}$ rotation around the $z$ axis in the previous works. 
Figures~\ref{fig:aa} (b) and (f) are the top view of \MS{} and \MSA{}, where the primitive cell is labeled by the parallelogram and show the similar lattices, except that \MSA{} is made up by hexagonal layers with A-B-A stacking (A: triangular pyramid; B: prism). 
{These figures also show that those two materials are related by a $\frac{\pi}{3}$ rotation along $z$-axis, and it is the reason why the Wyckoff positions of the S atom ($1b$) in the prism of \MS{} and the As atom ($1c$) in the prism of \MSA{} are different. Thus, care must be taken in comparing these materials and other TMD materials with different primitive cell conventions. }

Due to the honeycomb layer stacking structure, both \MSA{} and \MS{} will have valleys at $K$ ($\frac{1}{3},\frac{1}{3}$) and $K^{\prime}$ ($-\frac{1}{3},-\frac{1}{3}$) momenta in their electronic band structure. 
However, the difference of the primitive cell conventions in these two materials results 
%WPs for sulphur in \MS{} (with the unit-cell convention in previous works) and arsenic in \MSA{} forming the prism will make the prism in \MSA{} rotated by $\frac{\pi}{3}$ around $z$-axis with respect to the prism in \MS{}, resulting 
in a valley flipping between $K$ and $K^{\prime}$ in \MSA{}, in both the phonon and electronic band structures. 
Since $K$ and $K^{\prime}$ are time-reversal-related momenta, the valley $K$ in \MS{} will be equal to the valley $K^{\prime}$ in \MSA{}.

\subsection{Electronic band structures for \MSA{} and \MS}

Figures~\ref{fig:aa} (d) and (g) are the spinful electronic band structures for \MS{} and \MSA{}, respectively. 
Due to the similar prism crystal structures composed of Mo, \MSA{} and \MS{} also have similar band structures, i.e., valleys contributed by $d$ orbitals from Mo at $K$ and $K^{\prime}$. 
However, the $\frac{\pi}{3}$ rotation difference for the prisms in \MSA{} and \MS{} will also make an influence on their band structures, resulting in a valley flipping between two time-reversal-related momenta $K$ and $K^{\prime}$ for \MSA{}. 
%The reversal of two valleys can be also treated as different sign of the Ising SOC parameter at the same valley for \MSA{} and \MS{}, which will induce different spin-up and spin-down splitting for those two different materials at the same valley $K$. 
%Furthermore, \MS{} has an indirect gap due to the conduction band minimum (CBM) along $\Gamma-K$ direction, while \MSA{} has a direct one, which makes the hole doping not necessary for \MSA{} in the intervalley scattering process between $K$ and $K^{\prime}$. 
Since the rotation of the prism will also make the valence band maximum (VBM) orbital basis changed from $d_{x^2-y^2}-2id_{xy}$ at $K$ in \MS{} to $d_{x^2-y^2}+2id_{xy}$ at $K$ in \MSA{}, circular dichroism for the intervalley scattering process will also be different for those two materials. 
%All the differences between \MSA{} and \MS{} caused by the prism rotation will make an influence on the Bloch phase part of PAM for electrons, which will be discussed in detail later. 
In summary, \MSA{} has a direct gap at valleys $K$ and $K^{\prime}$ and larger SOC splitting, which are more preferable for valleytronics applications than \MS{}. 

{Since the conduction band minimum~(CBM) of \MSA{} is composed of 90\% $d_{z^2}$ and the VBM/(VMB-1) is composed of 85\% $d_{x^2-y^2}+2id_{xy}$ at $K$ valley, we can analyze the bands near the Fermi level for both TMDs and \MSA{} by pure orbital components. 
These atomic orbitals will contribute $l_{e}^o$ = 0 and +2 to the PAM for the valence band and conduction band, respectively. 
Therefore, the total electron PAM $l_e$ = $l_e^o$ + $l_e^s$ with spin components of $l_e^s=\pm \frac{1}{2}$ cannot be equal between different bands. 
%Furthermore, the spin parts with PAM of $l_{e}^s$ = $\pm\frac{1}{2}$ combines with $l_{e}^o$ = 0 and +2 will contribute to a total electron PAM of $l_{e}^s$ = $+\frac{1}{2}$, $-\frac{1}{2}$ and $+\frac{3}{2}$ for the CBM, VBM and VBM-1, separately. 
Thus, those bands cannot be hybridized with each other, resulting in a spin and orbital decoupled feature in both \MSA{} and TMDs~\cite{xiao2012coupled}. Further discussions are in Section II. B. }
%-------------------------------fig:cc, atomic motions ---------------------------%
\begin{center}
\begin{figure*}[htp]
\includegraphics[scale=0.8]{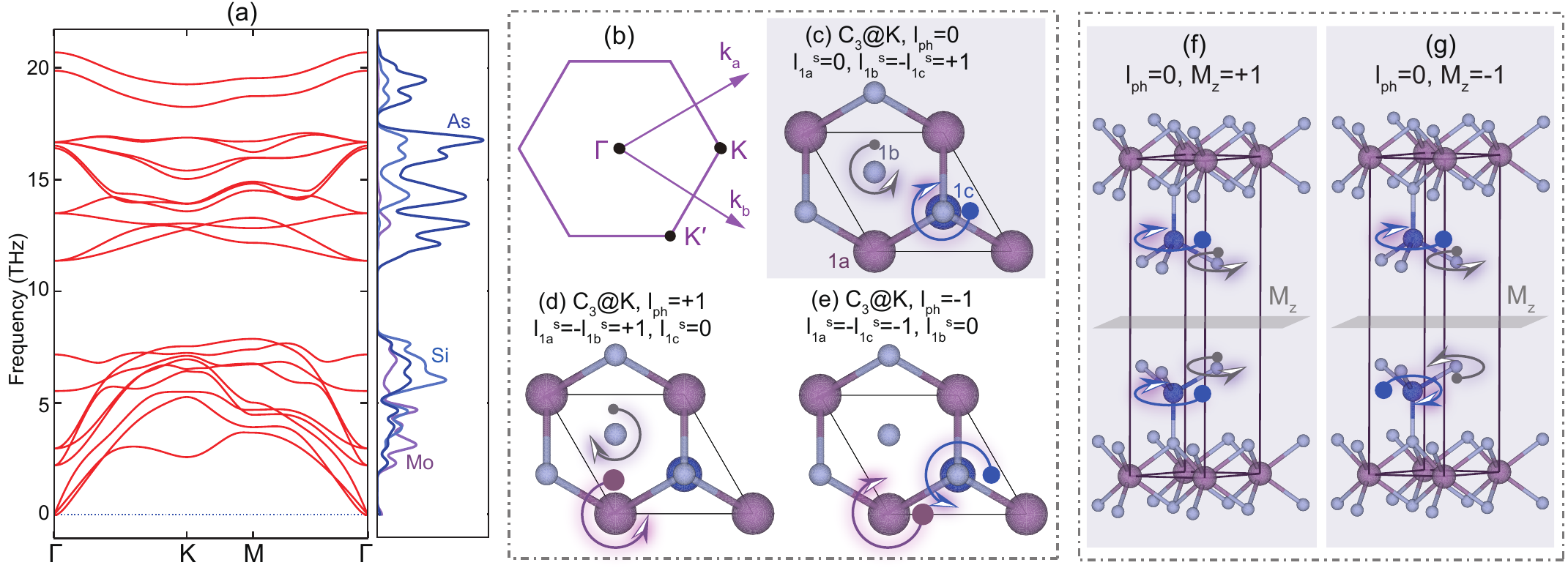}\caption{(a) Phonon spectra, density of states and (b) BZ for \MSA{}. (c-e) are the only three possible motions for atomic vibrations with $C_3$-invariant Wyckoff positons $1a$, $1b$ and $1c$ at $K$, which correspond to $l_{ph}=$ 0, $-1$ and +1, respectively. If $M_z$ symmetry is taken into consideration, each of the mode in (c-e) can be further cataloged into two cases: such as $l_{ph}=0$ in (b) can be cataloged into (f) with $M_z=+1$ and (g) with $M_z=-1$. We note that, in order to make the atomic motions for both silicon and arsenic at Wyckoff position $1c$ clear, we reduce the radius of arsenics in (c-g) to make silicons to be seen.
\label{fig:cc}}
\end{figure*}
\end{center}
%-----

\section{Circular dichroism for intervalley scattering}
Chiral phonons are collective excitations in crystals with nonzero circular polarization, which can be coupled with circularly polarized photons and valley electrons, resulting in chiral-phonon-involved circular dichroism. 
Large SOC, intrinsic spatial inversion symmetry breaking, and valleys at non-time-reversal-invariant momenta make TMDs ideal candidates to observe the chiral-phonon-involved circular dichroism at $K$ and $K^{\prime}$. 

%However, such process can be only realized between valence bands via hole doping in TMDs due to their indirect gaps.

In addition to all the advantages that TMDs have, \MSA{} is also a direct-gap semiconductor with larger SOC splitting, which makes \MSA{} a better platform for valleytronic and spintronic applications. 
Optical circular dichroism between conduction bands and valence bands without any doping can be also obtained in \MSA{} with the following conservations rules: 

\begin{equation}
l_{e}^{i}+m=l_{e}^{f}+l_{ph},
\end{equation}
\begin{equation}
\hbar\mathbf{k}_{e}^{i}=\hbar\mathbf{k}_{e}^{f}+\hbar\mathbf{q}_{ph},
\end{equation} 
\begin{equation}
E_{e}^{i}+E_{photon}=E_{e}^{f}+E_{ph},
\end{equation} 
where $l_{e}^{i/f}$ is the electron PAM for the initial/final state, $l_{ph}$ is the PAM for a phonon, $m$ = $\pm1$ represent the right/left-circularly polarized light, $\mathbf{k}_{e}$ and $\mathbf{q}_{ph}$ are the crystal momentum for electrons and phonons, $E_{e/photon/ph}$ is the energy for electrons/photons/phonons. 
In terms of the conservation rules above, we will introduce the definition and unify the conventions for each physical property involved in optical circular dichroism in the following, which are not clearly written and confusing in the previous studies. 

\subsection{Pseudo-angular momentum for phonons}
The phonon total PAM $l_{ph}$ is defined in terms of the eigenvalue of $n$-fold (screw) rotation symmetry:
\begin{equation}
\hat{C_{n}}u_{\mathbf{q}}=e^{\frac{-2\pi i}{n}\cdot l_{ph,\mathbf{q}}}u_{\mathbf{q}},
\label{eq:5}
\end{equation}
where $\hat{C_{n}}$ is the $n$-fold (screw) rotation operator, $u_{\mathbf{q}}$ is the phonon Bloch wave function, and $l_{ph,\mathbf{q}}$ is the phonon PAM at ${\mathbf{q}}$ with integer values of \{0, 1, ..., $n-1$\} $mod$ $n$ for symmorphic systems and with discrete values as a function of $\mathbf{q}$ for nonsymmorphic systems with screw rotation symmetry \cite{zhang2021chiral}. 

Since the phonon Bloch wave function $u_{\mathbf{q}}$ has both a nonlocal contribution (also called Bloch phase/intercell/orbital part $l_{ph}^{o}$) from the Bloch phase factor $e^{i\mathbf{R}_{l} \cdot \mathbf{q}}$ ($\mathbf{R}_{l}$ is the atomic WP) and a local contribution (also called self-rotation/intracell/spin part $l_{ph}^{s}$) from relative vibrations between sublattices, 
phonon PAM can be also decomposed into two terms: $l_{ph}=l_{ph}^{s}+l_{ph}^{o}$ \cite{yao2008valley}. 
$l_{ph}^{o}$ corresponds to the phase difference of the Bloch phase factor between atoms in different sublattices related by (screw) rotation symmetry, and it depends on the phonon momentum $\mathbf{q}$ and the WP $\mathbf{R}_{l}$. Like $l_{ph}$, $l_{ph}^{o}$ is also an integer for symmorphic systems, while it becomes a $\mathbf{q}$-dependent quantity in nonsymmorphic systems with screw rotation symmetry~\cite{zhang2021chiral}. 
For TMDs and \MSA{}, which have $C_{3z}$ symmetry ($n$=3), $l_{ph}^{o}\equiv 0$ for each atom with arbitrary phonon mode at $\Gamma$, while it will have different values of $l_{ph,1a}^{o}=0$, $l_{ph,1b}^{o}=+1$ and $l_{ph,1c}^{o}=-1$ at the $K$ valley, for the WPs $1a$, $1b$ and $1c$, respectively.

%%%--- comment
Next, we discuss the local contribution $l_{ph}^{s}$. At $\Gamma$, $l_{ph}^{s}$ is the phonon angular momentum (or atomic motion) in TMDs and \MSA{}. 
To discuss the value of $l_{ph}^{s}$ at valley $K$, we note that all the atoms are at $C_3$-invariant WPs in the crystals. 
Therefore, these crystals belong to Case I in Ref. \cite{zhang2021chiral}, meaning that there are limited possibilities for phonons at $K$. 
Since $l_{ph}^{o}$ only depends on the WPs, there is only one possibility for $l_{ph}^{s}$ at $K$ for each value of $l_{ph}$ ($=\pm1, 0$), resulting in three kinds of atomic motions for \MSA{} in total, as summarized in Figs.~\ref{fig:cc} (c-e). 

{To understand the phonon band structure, we also show the atomic projected density of states (DOS) in Fig.~\ref{fig:cc} (a). The DOS indicates that phonon modes with lower energies are mostly contributed by the vibrations of Mo and Si, while the ones with higher energies are mostly contributed by the vibrations of Si and As. 
As we discussed above, if we only take $C_3$ symmetry into consideration, all the phonon modes at $K$ only have three vibration modes, as shown in Figs.~\ref{fig:cc} (c-e). 
However, due to the additional horizontal mirror symmetry $M_z$ in \MSA{}, phonon modes with atomic vibrations shown in Figs.~\ref{fig:cc} (c-e) can be further divided into two possibilities with $M_z=+1$ and $M_z=-1$, which correspond to a 0 and $\pi$ phase difference for in-plane oscillations for two $M_z$-related atoms shown in Figs.~\ref{fig:cc} (f-g), respectively. }
Since physical process like intervalley scattering keeps even under $M_z$, only phonon modes with $M_z=+1$ can be involved. 
We note that there are 11 phonon modes with $M_z=+1$ at valley $K$, which contributes 5 more chiral phonon modes than that in \MS{}.

\subsection{Pseudo-angular momentum for electrons}
Like phonons, the total PAM for electrons $l_e$, which is calculated directly from the (screw) rotation operator, can be decomposed into the Bloch phase and self-rotation part $l_{e}=l_{e}^{o}+l_{e}^{s}$. 
$l_{e}^{s}$ is relatively simple for electrons, which is only related with the spin of a state. 
To meet the convention used in previous studies, we note that $l_{e}^{s}=-\frac{1}{2}$ for spin-up states, while $l_{e}^{s}=\frac{1}{2}$ for spin-down states. 
Meanwhile, $l_{e}^o$ should be paid more attention since it is the sum of the atomic orbital angular moment $l_{orb}$ for electrons and the interatomic part, which is identical with that for phonons, $l_{ph}^{o}$, determined by the atomic WPs. 

Particularly, since $l_{e}^o$=$l_{orb}$+$l_{ph}^{o}$, for two bands contributed from the same orbitals (equal value of $l_{orb}$) of the same atoms (equal value of $l_{ph}^{o}$), $l_{e}^o$ should be equal to each other. 
In Ref. \cite{chen2021propagating}, valence band maximum (VBM), and second valence band maximum (VBM$-1$) at valley $H$ for WN$_2$ are contributed by the same $d_{x^2-y^2}+2id_{xy}$ orbitals from tungsten (WP $1a$ with $l_{ph}^o$=0) but with opposite spin component according to our first-principle calculation. Thus $l_{e}^o$=$l_{orb}$+$l_{ph}^o$=$-2$+0=$-2$ for both VBM and (VBM$-1$) at valley $H$, as summarized in Tab.~\ref{table:wn2}, instead of taking the different values of $-1$ and 0 in Ref. \cite{chen2021propagating}. 
We note that, although WN$_2$ is a 3D material, it shares the same PAM definition with the ones defined in 2D systems. 
We also note that $l_{e}^s$ is omitted in Ref. \cite{chen2021propagating}, since it is already considered in the spin conservation rule. Thus, $l_{e}^o\equiv l_e$ in Ref. \cite{chen2021propagating}, which is also different with our results shown in Tab.~\ref{table:wn2}. 
However, $l_{e}^s$ is necessary to obtain $l_e$. 
Since the VBM and VBM$-1$ bands at $H$ are with spin-up and spin-down components, they will have $l_{e}^s=-\frac{1}{2}$ and $+\frac{1}{2}$, respectively, giving rise to $l_{e}=l_{e}^{o}+l_{e}^{s}=$ $+\frac{1}{2}$ and $+\frac{3}{2}$ in WN$_2$.

\begin{table}[]
\begin{tabular}{c|cccc|cccc}
\hline \hline
WN$_2$   & \multicolumn{4}{c|}{valley $H$}                                                                                    & \multicolumn{4}{c}{valley $H^{\prime}$}                                                               \\ \hline
items & $l_{e}^o$ & $l_{e}^s$   & \multicolumn{1}{c|}{$l_e$}   &\begin{tabular}[c]{@{}c@{}}$l_e$ in\\ Ref.~\cite{chen2021propagating}\end{tabular}  & $l_{e}^o$ & $l_{e}^s$   & \multicolumn{1}{c|}{$l_e$}   & \begin{tabular}[c]{@{}c@{}}$l_e$ in\\ Ref.~\cite{chen2021propagating}\end{tabular} \\ \hline
VBM   & $-2$ & $-\frac{1}{2}$ & \multicolumn{1}{l|}{$+\frac{1}{2}$} & 1   & +2 & $+\frac{1}{2}$ & \multicolumn{1}{c|}{$-\frac{1}{2}$} & $-1$                                                  \\
VBM-1 & $-2$ & $+\frac{1}{2}$ & \multicolumn{1}{l|}{$+\frac{3}{2}$} & 0  & +2 & $-\frac{1}{2}$ & \multicolumn{1}{c|}{$+\frac{3}{2}$} & 0                                                   \\ \hline
\end{tabular}
\caption{PAM calculation for two valence bands in WN$_2$. Since both of the VBM and VBM$-1$ bands are composed of $d_{x^2-y^2}-2id_{xy}$ orbitals from tungsten (WP $1a$), $l_{e}^o$ is given by $l_{e}^o$=$l_{orb}$+$l_{ph}^o$=$-2$+0=$-2$ at valley $H$ by following the unified convention in this paper. We note that the PAM in these materials are defined in terms of modulo 3. 
}
\label{table:wn2}
\end{table}

Under the unified convention for $l_{e}^o$, $l_{e}^s$ and $l_{e}$ discussed above, we calculate all the quantities for CBM, VBM and (VBM$-1$) in both \MS{} and \MSA{} at the $K$ valley, and the results are summarized in Tab.~\ref{table:pam}. 
For example, the VBM and VBM$-1$ of \MS{} are composed of $d_{x^2-y^2}-2id_{xy}$ orbital ($l_{orb}=-2$) from Mo atom, which has a WP of $1a$ ($l_{ph}^o$=0). Thus, $l_{e}^o$=$l_{orb}$+$l_{ph}^o$=$-2$+0=$-2$ for both of the valence bands. Furthermore, due to the different spin components for VBM and VBM$-1$, $l_{e}^s$ are also different, giving rise to $l_e$=$-2-\frac{1}{2}=-\frac{5}{2}$ ($mod$ 3) for the VBM and $l_e$=$-2+\frac{1}{2}=-\frac{3}{2}$ ($mod$ 3) for the VBM$-1$. 
As we discussed above, 
%the $K$ valley in \MSA{} will rotate $\frac{\pi}{3}$ and become $K^{\prime}$ comparing to the one in \MS{} 
due to the $\frac{\pi}{3}$ rotation of the prism in the real space, the $K$ valley in \MSA{} should be equal to the $K^{\prime}$ valley in \MS{}. 
Thus, since $K$ and $K^{\prime}$ are related by time-reversal symmetry for the same band, the $l_{e}^o$, $l_{e}^s$ and $l_{e}$ will have opposite signs between \MSA{} and \MS{}. 
For example, for the VBM in \MSA{}, $l_e^o$=$l_{orb}$+$l_{ph}^o$=+2+0=+2 due to the $d_{x^2-y^2}+2id_{xy}$ orbital contribution from Mo, as shown in Tab.~\ref{table:pam}. 
Since the $l_{e}^s$ for the VBM in \MSA{} and \MS{} will also have the opposite sign at $K$ valley, $l_e=l_e^o+l_e^s$=+2+$\frac{1}{2}$=$+\frac{5}{2}$=$-\frac{1}{2}$ ($mod$ 3), which has an opposite sign for the VBM with $l_e=+\frac{1}{2}$ in \MS{} as well. 
Since there is a valley flipping between $K$ and $K^{\prime}$ in \MSA{} comparing to TMDs, their valley-selective optical circular dichroism will also be different.

\begin{table}[]
\begin{tabular}{ccrrcc}
\hline 
material &band order   & $l_{e}^{o}$   & $l_{e}^s$ &$l_e$ calculated by & $l_{e}$ calculated by   \\ 
&at $K$ &&& $l_{e}^{s}+l_{e}^{o}$ &  eigenvalue of $\hat{C_{3}}$ \\ \hline\hline

&CBM  & 0                   & $-\frac{1}{2}$     &  $-\frac{1}{2}$ & $-\frac{1}{2}$ \\ 
\MS{} &VBM   & $-2$       & $-\frac{1}{2}$  &  $-\frac{5}{2}$($+\frac{1}{2}$)&  $+\frac{1}{2}$   \\ 
&VBM$-1$& $-2$            &  $+\frac{1}{2}$  &  $-\frac{3}{2}$   &  $-\frac{3}{2}$  \\ \hline

&CBM  & 0 & $+\frac{1}{2}$     &  $+\frac{1}{2}$ & $+\frac{1}{2}$ \\ 
\MSA{} &VBM   & +2 & $+\frac{1}{2}$  &  $+\frac{5}{2}$ ($-\frac{1}{2}$) &  $-\frac{1}{2}$   \\ 
&VBM$-1$& +2 & $-\frac{1}{2}$  &  $+\frac{3}{2}$ &  $+\frac{3}{2}$ \\ \hline
\end{tabular}
\caption{$l_{e}^o$, $l_e^{s}$ and $l_{e}$ ($mod$ 3) for the conduction band minimum (CBM), valence band maximum (VBM), and second valence band maximum (VBM$-1$) at valley $K$ ($\frac{1}{3},\frac{1}{3}$) in \MS{} and \MSA{}. 
Due to the $\frac{\pi}{3}$ rotation of the prism in crystal structures, the $K$ and $K^{\prime}$ valleys are exchanged between \MS{} and \MSA{}. Therefore, the value of $l_e^{o}$, $l_{e}^{s}$ and $l_{e}$ have opposite signs between \MS{} and \MSA{} for the same bands at the same valley $K$. } 
\label{table:pam}
\end{table}

{\subsection{Physical meanings of the nonlocal and local part of PAM for both phonons and electrons}}

{For phonons, the nonlocal part $l_{ph}^o$ corresponds to the phase difference of the Bloch phase factor between atoms related by the (screw) rotation symmetry. For electrons, in addition to the phase difference of the Bloch phase factor between atoms, which is equal to that of phonons (= $l_{ph}^o$), the atomic orbitals (such as $p$ and $d$ orbitals) also contribute to the nonlocal PAM. Thus, the orbital PAM of electrons should be the summation of $l_{ph}^o$ and its orbital angular momentum coming from atomic orbitals. In conclusion, $l_{ph}^o$ will contribute to the orbital PAM for both phonons and electrons, and it is originally from the atomic position information. 
} 

{
As for the local/spin part of PAM, phonons and electrons will also have different meanings. The local part of phonon PAM is contributed by the sub-lattice (relative) vibration, while the local part of electron PAM reflects the spin component of the band. 
}

\subsection{Valley-selective optical circular dichroism in \MSA{}}

%------------------------------- intervalley scattering ---------------------------%
\begin{center}
\begin{figure}[t]
\includegraphics[scale=0.8]{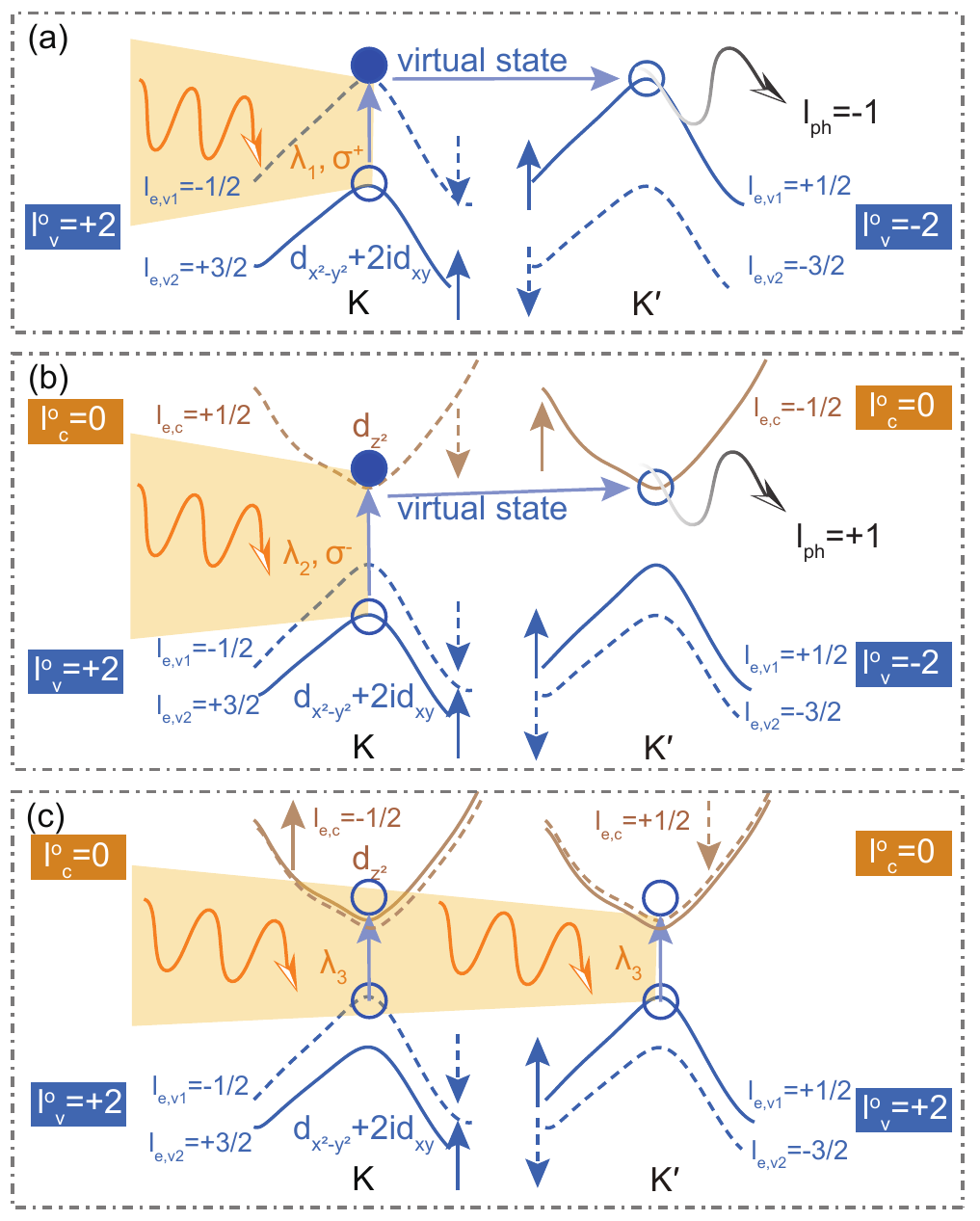}\caption{Selection rules for chiral phonon absorption/emission in the intervalley electron scattering process via circularly polarized photon absorption in \MSA. (a) Chiral phonons with $l_{ph}=-1$ emitted in \MSA{} with hole doping, where the hole at valley $K$ is excited by a right-hand circularly polarized photon $\sigma^{+}$ with energy $\lambda_{1}$. 
(b) Chiral phonon emission with $l_{ph}=+1$ via a left-hand circularly polarized photon $\sigma^{-}$ with energy $\lambda_{2}$, and this process does not need any doping. 
Likewise, excitations at valley $K^{\prime}$ can be selectively excited by $\sigma^{+}$, together with a chiral phonon emission with $l_{ph}=-1$. 
(c) Valley excitations at both valleys $K$ and $K^{\prime}$ via a non-polarized photon with energy $\lambda_{3}$, and there is no chiral phonon emission/absorption due to the direct excitation of electrons.  
\label{fig:bb}}
\end{figure}
\end{center}
%-------------------

{Chiral-phonon-involved optical circular dichroism were widely studied in the hole-doped \MS{} in the past, which can be also realized in \MSA{}. }
Figure~\ref{fig:bb} (a) shows the hole-doped intervalley scattering process in \MSA{} with electrons excited from VBM$-1$ at valley $K$ to VBM at valley $K^{\prime}$. 
Since the momentum difference for the electron and hole in the intervalley transition is $\mathbf{k}_{e,f} - \mathbf{k}_{e,i}$ = ($\frac{2}{3}$, $\frac{2}{3}$), a chiral phonon with momentum $\mathbf{q}_{ph}$ = ($-\frac{2}{3}$, $-\frac{2}{3}$) and $l_{ph}=-1$ will be emitted when the right-hand circularly polarized incident light ($\sigma^{+}$) with the energy $\lambda_1$ larger than the band gap between VBM$-1$ and VBM. 
Likewise, a chiral phonon with $l_{ph}=-1$ at $K$ will be emitted under the left-hand circularly polarized incident light ($\sigma^{-}$).

Besides the above case shared by both TMDs and \MSA{}, the optical circular dichroism can be also observed in \MSA{} without any doping. 
As shown in Fig.~\ref{fig:bb} (b), where the electrons are excited from {VBM-1} at valley $K$ to CBM at valley $K^{\prime}$ with chiral phonon $l_{ph}=+1$ emission for the incident $\sigma^{-}$ with energy $\lambda_2$ larger than the band gap between VBM and CBM. 
Likewise, electrons at valley $K^{\prime}$ can be selectively excited by the $\sigma^{+}$ with energy of $\lambda_2$, together with a chiral phonon emission with $l_{ph}=-1$, which is a time-reversal related process with the one shown in Fig.~\ref{fig:bb} (b). 
We note that both the chiral phonon emission and absorption processes with $l_{ph}=-1,+1$ can be obtained in \MSA{}, and they can be modulated by both of the energy and the handedness of the circularly polarized incident light. 

Valley excitations at both valleys $K$ and $K^{\prime}$ can be also obtained simultaneously in \MSA{}, which is triggered by a non-polarized photon with energy $\lambda_{3}$ ($\lambda_{3}$ is larger than the band gap between CBM and VBM), and there is no chiral phonon emission/absorption due to the direct excitation of electrons, as shown in Fig.~\ref{fig:bb} (c).  
%Optical photons do not carry significant momentum, so they cannot selectively populate different valleys based on this attribute. In certain materials, however, carriers in different valleys are associated with well-defined, but different angular momenta. This suggests the possibility of addressing different valleys by controlling the photon angular momentum, that is, by the helicity (circular polarization state) of light. 

%-------------- SHE ------------
\begin{center}
\begin{figure}[t]
\includegraphics[scale=0.5]{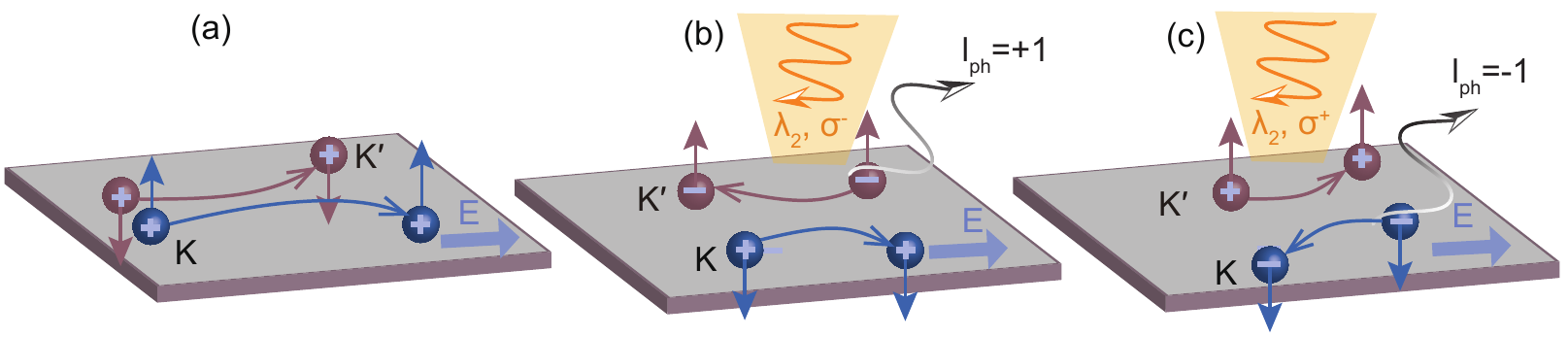}\caption{(a) Spin Hall effect and valley Hall effect for hole doped \MSA{} without charge Hall currents. The carriers at valleys $K$ and $K^{\prime}$ are marked in dark blue and dark purple, respectively. (b) Valley Hall effect with charge Hall currents under a left-hand circularly polarized light $\sigma^{-}$ with energy $\lambda_{2}$, and this process also involves a chiral phonon emission with $l_{ph}=+1$. (c) Valley Hall effect with opposite charge Hall currents under a right-hand circularly polarized light $\sigma^{+}$ with energy $\lambda_{2}$, and this process involves a chiral phonon emission with $l_{ph}=-1$. 
``+'' and ``$-$'' symbols represent for the hole and electron, up and down arrows represent for the spin-up and spin-down states, respectively. 
\label{fig:dd}}
\end{figure}
\end{center}
%-------------- SHE ------------

\section{Spin Hall effect and Valley Hall effect in \MSA{}}

Figure~\ref{fig:dd} (a) shows the spin Hall effect and valley Hall effect in \MSA{} with hole doping, where ``+'' represent holes at each valley. Since the Berry curvatures at valley $K$ and $K^{\prime}$ have different signs, holes at different valleys will move in opposite directions carrying opposite anomalous transverse velocities, which is similar with \MS{}~\cite{xiao2012coupled}. 
{Spin Hall effect and valley Hall effect can be also obtained in \MSA{} without any doping, which is also accompanied by a chiral phonon emission/absorption process. }

Figure~\ref{fig:dd} (b) shows multiple Hall effects, i.e., spin Hall effect and valley Hall effect, triggered by a left-hand circularly polarized light $\sigma^{-}$, which corresponds to the intervalley scattering process in Fig.~\ref{fig:bb} (b). 
``+'' and ``$-$'' represent holes and electrons at different valleys. 
If the energy of the incident light $\lambda_2$ is slightly larger (smaller) than the energy gap between VBM$-1$ and CBM at $K$, a chiral phonon $l_{ph}=+1$ emission (absorption) will be also triggered, together with the multiple Hall effects. 
Since electrons and holes at different valleys will carry transverse velocities with different sign in both cases, they will be accumulated on different boundaries, leading to a transverse Hall voltage. 
{We note that the spin and valley Hall effects discussed in Ref.~\cite{xiao2012coupled} are intravalley scattering processes, in which both the electrons and holes accumulated at the boundaries are from the same valley. However, in Fig.~\ref{fig:dd} (b), the electrons and holes accumulated at the boundaries are from different valleys, due to the phonon-involved intervalley scattering process. }

Figure~\ref{fig:dd} (c) shows a time-reversed process with Fig.~\ref{fig:dd} (b), which will have a transverse voltage and chiral phonon emission/absorption with opposite sign. 
We conclude that the spin, valley, chiral phonon as well as Hall voltage can be all modulated by altering the handedness and the energy of the circularly polarized incident light, which makes \MSA{} an ideal platform for the valleytronics and logic gate applications.

\paragraph*{Conclusion}
%Thus, spin and valley are coupled by external light, which are similar with other 2D TMDs. 
%
As a new family of 2D materials, $X$Si$_2Y_4$ ($X$: transition metals; $Y$: nitrogen group elements) are ideal platforms for practical applications. 
By comparing the similarities and differences between TMDs and $X$Si$_2Y_4$, with \MS{} and \MSA{} as examples, we unify the convention of the definitions for both the phonon PAM and the electron PAM. 
Besides all the physical properties that TMDs have,  the chiral phonon emission and absorption without any doping can be also obtained in \MSA{}, companied by multiple Hall effects at the same time. 
Thus, the spin, the valley and the chiral phonon degrees of freedom in \MSA{} are coupled with external circularly polarized light, including both the energy and chirality.  
Furthermore, transverse Hall voltage can be a hallmark for both the chiral phonon emission/absorption process and modulating the spin and the valley degrees of freedom for \MSA{}, which makes \MSA{} a better platform for practical applications like spintronics and logic gate devices.

\section*{methods}
First-principles calculations on both the spinful electronic bands and phonon bands are implemented by Vienna $ab$ $initio$ simulation package~\cite{CAL_VASP} with the generalized gradient approximation~\cite{DFT_GGA} and an energy cutoff of 302 eV. 
A $9\times9\times1$ $k$-mesh was used in the BZ for the self-consistent calculations on electronic band structures, while an equivalent $k$-mesh of $8\times8\times1$ was used for phonon calculation. 
Prior to the phonon calculations, crystal structure of \MSA{} was relaxed until the residual force on each atom less than 0.01 $eV/\AA$. 
A vacuum space of 25 $\AA$ perpendicular to the monolayer plane is applied to separate the periodic images, and weak vdW interaction between adjacent monolayers is described by the DFT-D3 functional with Grimme correction~\cite{grimme2010consistent,grimme2011effect}.

\paragraph*{Acknowledgement}
We acknowledge the supports from Tokodai Institute for Element Strategy (TIES) funded by MEXT Elements Strategy Initiative to Form Core Research Center Grants No. JPMXP0112101001, JP18J23289, JP18H03678, and JP22H00108. T. Z. also acknowledge the support by Japan Society for the Promotion of Science (JSPS), KAKENHI Grant No. JP21K13865.

%
%
%\paragraph*{Competing interets}
%The authors declare no competing interests.
%
%\paragraph*{Author contributions}
%All authors contributed to the performance of the project and to the discussion and the writing of the paper.
%
%\paragraph*{Data Availability}
%The data that support the findings of this study are available from the corresponding author upon reasonable request.

%\bibliographystyle{unsrt}
\bibliography{reference}

\begin{thebibliography}{51}%
\makeatletter
\providecommand \@ifxundefined [1]{%
 \@ifx{#1\undefined}
}%
\providecommand \@ifnum [1]{%
 \ifnum #1\expandafter \@firstoftwo
 \else \expandafter \@secondoftwo
 \fi
}%
\providecommand \@ifx [1]{%
 \ifx #1\expandafter \@firstoftwo
 \else \expandafter \@secondoftwo
 \fi
}%
\providecommand \natexlab [1]{#1}%
\providecommand \enquote  [1]{``#1''}%
\providecommand \bibnamefont  [1]{#1}%
\providecommand \bibfnamefont [1]{#1}%
\providecommand \citenamefont [1]{#1}%
\providecommand \href@noop [0]{\@secondoftwo}%
\providecommand \href [0]{\begingroup \@sanitize@url \@href}%
\providecommand \@href[1]{\@@startlink{#1}\@@href}%
\providecommand \@@href[1]{\endgroup#1\@@endlink}%
\providecommand \@sanitize@url [0]{\catcode `\\12\catcode `\$12\catcode
  `\&12\catcode `\#12\catcode `\^12\catcode `\_12\catcode `\%12\relax}%
\providecommand \@@startlink[1]{}%
\providecommand \@@endlink[0]{}%
\providecommand \url  [0]{\begingroup\@sanitize@url \@url }%
\providecommand \@url [1]{\endgroup\@href {#1}{\urlprefix }}%
\providecommand \urlprefix  [0]{URL }%
\providecommand \Eprint [0]{\href }%
\providecommand \doibase [0]{http://dx.doi.org/}%
\providecommand \selectlanguage [0]{\@gobble}%
\providecommand \bibinfo  [0]{\@secondoftwo}%
\providecommand \bibfield  [0]{\@secondoftwo}%
\providecommand \translation [1]{[#1]}%
\providecommand \BibitemOpen [0]{}%
\providecommand \bibitemStop [0]{}%
\providecommand \bibitemNoStop [0]{.\EOS\space}%
\providecommand \EOS [0]{\spacefactor3000\relax}%
\providecommand \BibitemShut  [1]{\csname bibitem#1\endcsname}%
\let\auto@bib@innerbib\@empty
%</preamble>
\bibitem [{\citenamefont {Feldman}\ \emph {et~al.}(1995)\citenamefont
  {Feldman}, \citenamefont {Wasserman}, \citenamefont {Srolovitz},\ and\
  \citenamefont {Tenne}}]{feldman1995high}%
  \BibitemOpen
  \bibfield  {author} {\bibinfo {author} {\bibfnamefont {Y.}~\bibnamefont
  {Feldman}}, \bibinfo {author} {\bibfnamefont {E.}~\bibnamefont {Wasserman}},
  \bibinfo {author} {\bibfnamefont {D.}~\bibnamefont {Srolovitz}}, \ and\
  \bibinfo {author} {\bibfnamefont {R.}~\bibnamefont {Tenne}},\ }\href@noop {}
  {\bibfield  {journal} {\bibinfo  {journal} {Science}\ }\textbf {\bibinfo
  {volume} {267}},\ \bibinfo {pages} {222} (\bibinfo {year}
  {1995})}\BibitemShut {NoStop}%
\bibitem [{\citenamefont {Wang}\ \emph {et~al.}(2015)\citenamefont {Wang},
  \citenamefont {Wang},\ and\ \citenamefont {Warner}}]{wang2015all}%
  \BibitemOpen
  \bibfield  {author} {\bibinfo {author} {\bibfnamefont {S.}~\bibnamefont
  {Wang}}, \bibinfo {author} {\bibfnamefont {X.}~\bibnamefont {Wang}}, \ and\
  \bibinfo {author} {\bibfnamefont {J.~H.}\ \bibnamefont {Warner}},\
  }\href@noop {} {\bibfield  {journal} {\bibinfo  {journal} {{ACS} nano}\
  }\textbf {\bibinfo {volume} {9}},\ \bibinfo {pages} {5246} (\bibinfo {year}
  {2015})}\BibitemShut {NoStop}%
\bibitem [{\citenamefont {Chang}\ \emph {et~al.}(2020)\citenamefont {Chang},
  \citenamefont {Ho}, \citenamefont {Tseng}, \citenamefont {Lin}, \citenamefont
  {Hou}, \citenamefont {Lin}, \citenamefont {Wang}, \citenamefont {Huang},
  \citenamefont {Chiang}, \citenamefont {Yang} \emph {et~al.}}]{chang2020fast}%
  \BibitemOpen
  \bibfield  {author} {\bibinfo {author} {\bibfnamefont {M.-C.}\ \bibnamefont
  {Chang}}, \bibinfo {author} {\bibfnamefont {P.-H.}\ \bibnamefont {Ho}},
  \bibinfo {author} {\bibfnamefont {M.-F.}\ \bibnamefont {Tseng}}, \bibinfo
  {author} {\bibfnamefont {F.-Y.}\ \bibnamefont {Lin}}, \bibinfo {author}
  {\bibfnamefont {C.-H.}\ \bibnamefont {Hou}}, \bibinfo {author} {\bibfnamefont
  {I.-K.}\ \bibnamefont {Lin}}, \bibinfo {author} {\bibfnamefont
  {H.}~\bibnamefont {Wang}}, \bibinfo {author} {\bibfnamefont {P.-P.}\
  \bibnamefont {Huang}}, \bibinfo {author} {\bibfnamefont {C.-H.}\ \bibnamefont
  {Chiang}}, \bibinfo {author} {\bibfnamefont {Y.-C.}\ \bibnamefont {Yang}},
  \emph {et~al.},\ }\href@noop {} {\bibfield  {journal} {\bibinfo  {journal}
  {Nature communications}\ }\textbf {\bibinfo {volume} {11}},\ \bibinfo {pages}
  {1} (\bibinfo {year} {2020})}\BibitemShut {NoStop}%
\bibitem [{\citenamefont {Radisavljevic}\ \emph {et~al.}(2011)\citenamefont
  {Radisavljevic}, \citenamefont {Radenovic}, \citenamefont {Brivio},
  \citenamefont {Giacometti},\ and\ \citenamefont
  {Kis}}]{radisavljevic2011single}%
  \BibitemOpen
  \bibfield  {author} {\bibinfo {author} {\bibfnamefont {B.}~\bibnamefont
  {Radisavljevic}}, \bibinfo {author} {\bibfnamefont {A.}~\bibnamefont
  {Radenovic}}, \bibinfo {author} {\bibfnamefont {J.}~\bibnamefont {Brivio}},
  \bibinfo {author} {\bibfnamefont {V.}~\bibnamefont {Giacometti}}, \ and\
  \bibinfo {author} {\bibfnamefont {A.}~\bibnamefont {Kis}},\ }\href@noop {}
  {\bibfield  {journal} {\bibinfo  {journal} {Nature nanotechnology}\ }\textbf
  {\bibinfo {volume} {6}},\ \bibinfo {pages} {147} (\bibinfo {year}
  {2011})}\BibitemShut {NoStop}%
\bibitem [{\citenamefont {Hong}\ \emph {et~al.}(2020)\citenamefont {Hong},
  \citenamefont {Liu}, \citenamefont {Wang}, \citenamefont {Zhou},
  \citenamefont {Ma}, \citenamefont {Xu}, \citenamefont {Feng}, \citenamefont
  {Chen}, \citenamefont {Chen}, \citenamefont {Sun}, \citenamefont {Chen},
  \citenamefont {Cheng},\ and\ \citenamefont {Ren}}]{Hong2020}%
  \BibitemOpen
  \bibfield  {author} {\bibinfo {author} {\bibfnamefont {Y.-L.}\ \bibnamefont
  {Hong}}, \bibinfo {author} {\bibfnamefont {Z.}~\bibnamefont {Liu}}, \bibinfo
  {author} {\bibfnamefont {L.}~\bibnamefont {Wang}}, \bibinfo {author}
  {\bibfnamefont {T.}~\bibnamefont {Zhou}}, \bibinfo {author} {\bibfnamefont
  {W.}~\bibnamefont {Ma}}, \bibinfo {author} {\bibfnamefont {C.}~\bibnamefont
  {Xu}}, \bibinfo {author} {\bibfnamefont {S.}~\bibnamefont {Feng}}, \bibinfo
  {author} {\bibfnamefont {L.}~\bibnamefont {Chen}}, \bibinfo {author}
  {\bibfnamefont {M.-L.}\ \bibnamefont {Chen}}, \bibinfo {author}
  {\bibfnamefont {D.-M.}\ \bibnamefont {Sun}}, \bibinfo {author} {\bibfnamefont
  {X.-Q.}\ \bibnamefont {Chen}}, \bibinfo {author} {\bibfnamefont {H.-M.}\
  \bibnamefont {Cheng}}, \ and\ \bibinfo {author} {\bibfnamefont
  {W.}~\bibnamefont {Ren}},\ }\href {\doibase 10.1126/SCIENCE.ABB7023}
  {\bibfield  {journal} {\bibinfo  {journal} {Science}\ }\textbf {\bibinfo
  {volume} {369}},\ \bibinfo {pages} {670} (\bibinfo {year}
  {2020})}\BibitemShut {NoStop}%
\bibitem [{\citenamefont {Novoselov}(2020)}]{Novoselov2020}%
  \BibitemOpen
  \bibfield  {author} {\bibinfo {author} {\bibfnamefont {K.~S.}\ \bibnamefont
  {Novoselov}},\ }\href {\doibase 10.1093/NSR/NWAA190} {\bibfield  {journal}
  {\bibinfo  {journal} {National Science Review}\ }\textbf {\bibinfo {volume}
  {7}},\ \bibinfo {pages} {1842} (\bibinfo {year} {2020})}\BibitemShut
  {NoStop}%
\bibitem [{\citenamefont {Li}\ \emph {et~al.}(2020)\citenamefont {Li},
  \citenamefont {Wu}, \citenamefont {Feng}, \citenamefont {Guan}, \citenamefont
  {Feng}, \citenamefont {Yao},\ and\ \citenamefont {Yang}}]{li2020valley}%
  \BibitemOpen
  \bibfield  {author} {\bibinfo {author} {\bibfnamefont {S.}~\bibnamefont
  {Li}}, \bibinfo {author} {\bibfnamefont {W.}~\bibnamefont {Wu}}, \bibinfo
  {author} {\bibfnamefont {X.}~\bibnamefont {Feng}}, \bibinfo {author}
  {\bibfnamefont {S.}~\bibnamefont {Guan}}, \bibinfo {author} {\bibfnamefont
  {W.}~\bibnamefont {Feng}}, \bibinfo {author} {\bibfnamefont {Y.}~\bibnamefont
  {Yao}}, \ and\ \bibinfo {author} {\bibfnamefont {S.~A.}\ \bibnamefont
  {Yang}},\ }\href@noop {} {\bibfield  {journal} {\bibinfo  {journal} {Physical
  {R}eview {B}}\ }\textbf {\bibinfo {volume} {102}},\ \bibinfo {pages} {235435}
  (\bibinfo {year} {2020})}\BibitemShut {NoStop}%
\bibitem [{\citenamefont {Chen}\ \emph
  {et~al.}(2021{\natexlab{a}})\citenamefont {Chen}, \citenamefont {Chen},\ and\
  \citenamefont {Zhang}}]{chen2021first}%
  \BibitemOpen
  \bibfield  {author} {\bibinfo {author} {\bibfnamefont {R.}~\bibnamefont
  {Chen}}, \bibinfo {author} {\bibfnamefont {D.}~\bibnamefont {Chen}}, \ and\
  \bibinfo {author} {\bibfnamefont {W.}~\bibnamefont {Zhang}},\ }\href@noop {}
  {\bibfield  {journal} {\bibinfo  {journal} {Results in Physics}\ }\textbf
  {\bibinfo {volume} {30}},\ \bibinfo {pages} {104864} (\bibinfo {year}
  {2021}{\natexlab{a}})}\BibitemShut {NoStop}%
\bibitem [{\citenamefont {Yao}\ \emph {et~al.}(2021)\citenamefont {Yao},
  \citenamefont {Zhang}, \citenamefont {Wang}, \citenamefont {Li},
  \citenamefont {Yu}, \citenamefont {Xu}, \citenamefont {Wang},\ and\
  \citenamefont {Wei}}]{Yao2021}%
  \BibitemOpen
  \bibfield  {author} {\bibinfo {author} {\bibfnamefont {H.}~\bibnamefont
  {Yao}}, \bibinfo {author} {\bibfnamefont {C.}~\bibnamefont {Zhang}}, \bibinfo
  {author} {\bibfnamefont {Q.}~\bibnamefont {Wang}}, \bibinfo {author}
  {\bibfnamefont {J.}~\bibnamefont {Li}}, \bibinfo {author} {\bibfnamefont
  {Y.}~\bibnamefont {Yu}}, \bibinfo {author} {\bibfnamefont {F.}~\bibnamefont
  {Xu}}, \bibinfo {author} {\bibfnamefont {B.}~\bibnamefont {Wang}}, \ and\
  \bibinfo {author} {\bibfnamefont {Y.}~\bibnamefont {Wei}},\ }\href {\doibase
  10.3390/NANO11030559} {\bibfield  {journal} {\bibinfo  {journal}
  {Nanomaterials}\ }\textbf {\bibinfo {volume} {11}},\ \bibinfo {pages} {559}
  (\bibinfo {year} {2021})}\BibitemShut {NoStop}%
\bibitem [{\citenamefont {Yang}\ \emph {et~al.}(2021)\citenamefont {Yang},
  \citenamefont {Zhao}, \citenamefont {Shi-Qi}, \citenamefont {Liu},
  \citenamefont {Wang}, \citenamefont {Chen}, \citenamefont {Gao},\ and\
  \citenamefont {Zhao}}]{yang2021accurate}%
  \BibitemOpen
  \bibfield  {author} {\bibinfo {author} {\bibfnamefont {J.-S.}\ \bibnamefont
  {Yang}}, \bibinfo {author} {\bibfnamefont {L.}~\bibnamefont {Zhao}}, \bibinfo
  {author} {\bibfnamefont {L.}~\bibnamefont {Shi-Qi}}, \bibinfo {author}
  {\bibfnamefont {H.}~\bibnamefont {Liu}}, \bibinfo {author} {\bibfnamefont
  {L.}~\bibnamefont {Wang}}, \bibinfo {author} {\bibfnamefont {M.}~\bibnamefont
  {Chen}}, \bibinfo {author} {\bibfnamefont {J.}~\bibnamefont {Gao}}, \ and\
  \bibinfo {author} {\bibfnamefont {J.}~\bibnamefont {Zhao}},\ }\href@noop {}
  {\bibfield  {journal} {\bibinfo  {journal} {Nanoscale}\ }\textbf {\bibinfo
  {volume} {13}},\ \bibinfo {pages} {5479} (\bibinfo {year}
  {2021})}\BibitemShut {NoStop}%
\bibitem [{\citenamefont {Wu}\ \emph {et~al.}(2021)\citenamefont {Wu},
  \citenamefont {Cao}, \citenamefont {Ang},\ and\ \citenamefont
  {Ang}}]{wu2021semiconductor}%
  \BibitemOpen
  \bibfield  {author} {\bibinfo {author} {\bibfnamefont {Q.}~\bibnamefont
  {Wu}}, \bibinfo {author} {\bibfnamefont {L.}~\bibnamefont {Cao}}, \bibinfo
  {author} {\bibfnamefont {Y.~S.}\ \bibnamefont {Ang}}, \ and\ \bibinfo
  {author} {\bibfnamefont {L.~K.}\ \bibnamefont {Ang}},\ }\href@noop {}
  {\bibfield  {journal} {\bibinfo  {journal} {Applied Physics Letters}\
  }\textbf {\bibinfo {volume} {118}},\ \bibinfo {pages} {113102} (\bibinfo
  {year} {2021})}\BibitemShut {NoStop}%
\bibitem [{\citenamefont {Zhong}\ \emph {et~al.}(2021)\citenamefont {Zhong},
  \citenamefont {Xiong}, \citenamefont {Lv}, \citenamefont {Yu},\ and\
  \citenamefont {Yuan}}]{zhong2021strain}%
  \BibitemOpen
  \bibfield  {author} {\bibinfo {author} {\bibfnamefont {H.}~\bibnamefont
  {Zhong}}, \bibinfo {author} {\bibfnamefont {W.}~\bibnamefont {Xiong}},
  \bibinfo {author} {\bibfnamefont {P.}~\bibnamefont {Lv}}, \bibinfo {author}
  {\bibfnamefont {J.}~\bibnamefont {Yu}}, \ and\ \bibinfo {author}
  {\bibfnamefont {S.}~\bibnamefont {Yuan}},\ }\href {\doibase
  10.1103/PhysRevB.103.085124} {\bibfield  {journal} {\bibinfo  {journal}
  {Phys. Rev. B}\ }\textbf {\bibinfo {volume} {103}},\ \bibinfo {pages}
  {085124} (\bibinfo {year} {2021})}\BibitemShut {NoStop}%
\bibitem [{\citenamefont {Bafekry}\ \emph {et~al.}(2021)\citenamefont
  {Bafekry}, \citenamefont {Faraji}, \citenamefont {Hoat}, \citenamefont
  {Shahrokhi}, \citenamefont {Fadlallah}, \citenamefont {Shojaei},
  \citenamefont {Feghhi}, \citenamefont {Ghergherehchi},\ and\ \citenamefont
  {Gogova}}]{Bafekry_2021}%
  \BibitemOpen
  \bibfield  {author} {\bibinfo {author} {\bibfnamefont {A.}~\bibnamefont
  {Bafekry}}, \bibinfo {author} {\bibfnamefont {M.}~\bibnamefont {Faraji}},
  \bibinfo {author} {\bibfnamefont {D.~M.}\ \bibnamefont {Hoat}}, \bibinfo
  {author} {\bibfnamefont {M.}~\bibnamefont {Shahrokhi}}, \bibinfo {author}
  {\bibfnamefont {M.~M.}\ \bibnamefont {Fadlallah}}, \bibinfo {author}
  {\bibfnamefont {F.}~\bibnamefont {Shojaei}}, \bibinfo {author} {\bibfnamefont
  {S.~A.~H.}\ \bibnamefont {Feghhi}}, \bibinfo {author} {\bibfnamefont
  {M.}~\bibnamefont {Ghergherehchi}}, \ and\ \bibinfo {author} {\bibfnamefont
  {D.}~\bibnamefont {Gogova}},\ }\href {\doibase 10.1088/1361-6463/abdb6b} {\
  \textbf {\bibinfo {volume} {54}},\ \bibinfo {pages} {155303} (\bibinfo {year}
  {2021})}\BibitemShut {NoStop}%
\bibitem [{\citenamefont {Kang}\ and\ \citenamefont
  {Lin}(2021)}]{kang2021second}%
  \BibitemOpen
  \bibfield  {author} {\bibinfo {author} {\bibfnamefont {L.}~\bibnamefont
  {Kang}}\ and\ \bibinfo {author} {\bibfnamefont {Z.}~\bibnamefont {Lin}},\
  }\href {\doibase 10.1103/PhysRevB.103.195404} {\bibfield  {journal} {\bibinfo
   {journal} {Phys. Rev. B}\ }\textbf {\bibinfo {volume} {103}},\ \bibinfo
  {pages} {195404} (\bibinfo {year} {2021})}\BibitemShut {NoStop}%
\bibitem [{\citenamefont {Cui}\ \emph {et~al.}(2021)\citenamefont {Cui},
  \citenamefont {Luo}, \citenamefont {Yu},\ and\ \citenamefont
  {Xu}}]{cui2021tuning}%
  \BibitemOpen
  \bibfield  {author} {\bibinfo {author} {\bibfnamefont {Z.}~\bibnamefont
  {Cui}}, \bibinfo {author} {\bibfnamefont {Y.}~\bibnamefont {Luo}}, \bibinfo
  {author} {\bibfnamefont {J.}~\bibnamefont {Yu}}, \ and\ \bibinfo {author}
  {\bibfnamefont {Y.}~\bibnamefont {Xu}},\ }\href@noop {} {\bibfield  {journal}
  {\bibinfo  {journal} {Physica E: Low-dimensional Systems and Nanostructures}\
  }\textbf {\bibinfo {volume} {134}},\ \bibinfo {pages} {114873} (\bibinfo
  {year} {2021})}\BibitemShut {NoStop}%
\bibitem [{\citenamefont {Guo}\ \emph {et~al.}(2020)\citenamefont {Guo},
  \citenamefont {Zhu}, \citenamefont {Mu},\ and\ \citenamefont
  {Ren}}]{guo2020intrinsic}%
  \BibitemOpen
  \bibfield  {author} {\bibinfo {author} {\bibfnamefont {S.-D.}\ \bibnamefont
  {Guo}}, \bibinfo {author} {\bibfnamefont {Y.-T.}\ \bibnamefont {Zhu}},
  \bibinfo {author} {\bibfnamefont {W.-Q.}\ \bibnamefont {Mu}}, \ and\ \bibinfo
  {author} {\bibfnamefont {W.-C.}\ \bibnamefont {Ren}},\ }\href@noop {}
  {\bibfield  {journal} {\bibinfo  {journal} {EPL (Europhysics Letters)}\
  }\textbf {\bibinfo {volume} {132}},\ \bibinfo {pages} {57002} (\bibinfo
  {year} {2020})}\BibitemShut {NoStop}%
\bibitem [{\citenamefont {Mortazavi}\ \emph {et~al.}(2021)\citenamefont
  {Mortazavi}, \citenamefont {Javvaji}, \citenamefont {Shojaei}, \citenamefont
  {Rabczuk}, \citenamefont {Shapeev},\ and\ \citenamefont
  {Zhuang}}]{mortazavi2021exceptional}%
  \BibitemOpen
  \bibfield  {author} {\bibinfo {author} {\bibfnamefont {B.}~\bibnamefont
  {Mortazavi}}, \bibinfo {author} {\bibfnamefont {B.}~\bibnamefont {Javvaji}},
  \bibinfo {author} {\bibfnamefont {F.}~\bibnamefont {Shojaei}}, \bibinfo
  {author} {\bibfnamefont {T.}~\bibnamefont {Rabczuk}}, \bibinfo {author}
  {\bibfnamefont {A.~V.}\ \bibnamefont {Shapeev}}, \ and\ \bibinfo {author}
  {\bibfnamefont {X.}~\bibnamefont {Zhuang}},\ }\href@noop {} {\bibfield
  {journal} {\bibinfo  {journal} {Nano Energy}\ }\textbf {\bibinfo {volume}
  {82}},\ \bibinfo {pages} {105716} (\bibinfo {year} {2021})}\BibitemShut
  {NoStop}%
\bibitem [{\citenamefont {Guo}\ \emph {et~al.}(2021)\citenamefont {Guo},
  \citenamefont {Zhu}, \citenamefont {Mu}, \citenamefont {Wang},\ and\
  \citenamefont {Chen}}]{GUO2021110223}%
  \BibitemOpen
  \bibfield  {author} {\bibinfo {author} {\bibfnamefont {S.-D.}\ \bibnamefont
  {Guo}}, \bibinfo {author} {\bibfnamefont {Y.-T.}\ \bibnamefont {Zhu}},
  \bibinfo {author} {\bibfnamefont {W.-Q.}\ \bibnamefont {Mu}}, \bibinfo
  {author} {\bibfnamefont {L.}~\bibnamefont {Wang}}, \ and\ \bibinfo {author}
  {\bibfnamefont {X.-Q.}\ \bibnamefont {Chen}},\ }\href {\doibase
  https://doi.org/10.1016/j.commatsci.2020.110223} {\bibfield  {journal}
  {\bibinfo  {journal} {Computational Materials Science}\ }\textbf {\bibinfo
  {volume} {188}},\ \bibinfo {pages} {110223} (\bibinfo {year}
  {2021})}\BibitemShut {NoStop}%
\bibitem [{\citenamefont {Li}\ \emph {et~al.}(2021)\citenamefont {Li},
  \citenamefont {Zhou}, \citenamefont {Wan},\ and\ \citenamefont
  {Zhou}}]{LI2021114753}%
  \BibitemOpen
  \bibfield  {author} {\bibinfo {author} {\bibfnamefont {Q.}~\bibnamefont
  {Li}}, \bibinfo {author} {\bibfnamefont {W.}~\bibnamefont {Zhou}}, \bibinfo
  {author} {\bibfnamefont {X.}~\bibnamefont {Wan}}, \ and\ \bibinfo {author}
  {\bibfnamefont {J.}~\bibnamefont {Zhou}},\ }\href {\doibase
  https://doi.org/10.1016/j.physe.2021.114753} {\bibfield  {journal} {\bibinfo
  {journal} {Physica E: Low-dimensional Systems and Nanostructures}\ }\textbf
  {\bibinfo {volume} {131}},\ \bibinfo {pages} {114753} (\bibinfo {year}
  {2021})}\BibitemShut {NoStop}%
\bibitem [{\citenamefont {Wang}\ \emph
  {et~al.}(2021{\natexlab{a}})\citenamefont {Wang}, \citenamefont {Cao},
  \citenamefont {Liang}, \citenamefont {Wu}, \citenamefont {Wang},
  \citenamefont {Lee}, \citenamefont {Ong}, \citenamefont {Yang}, \citenamefont
  {Ang}, \citenamefont {Yang},\ and\ \citenamefont {Ang}}]{wang2021article}%
  \BibitemOpen
  \bibfield  {author} {\bibinfo {author} {\bibfnamefont {Q.}~\bibnamefont
  {Wang}}, \bibinfo {author} {\bibfnamefont {L.}~\bibnamefont {Cao}}, \bibinfo
  {author} {\bibfnamefont {S.-J.}\ \bibnamefont {Liang}}, \bibinfo {author}
  {\bibfnamefont {W.}~\bibnamefont {Wu}}, \bibinfo {author} {\bibfnamefont
  {G.}~\bibnamefont {Wang}}, \bibinfo {author} {\bibfnamefont {C.~H.}\
  \bibnamefont {Lee}}, \bibinfo {author} {\bibfnamefont {W.~L.}\ \bibnamefont
  {Ong}}, \bibinfo {author} {\bibfnamefont {H.~Y.}\ \bibnamefont {Yang}},
  \bibinfo {author} {\bibfnamefont {L.~K.}\ \bibnamefont {Ang}}, \bibinfo
  {author} {\bibfnamefont {S.~A.}\ \bibnamefont {Yang}}, \ and\ \bibinfo
  {author} {\bibfnamefont {Y.~S.}\ \bibnamefont {Ang}},\ }\href {\doibase
  10.1038/s41699-021-00251-y} {\bibfield  {journal} {\bibinfo  {journal} {npj
  2D Materials and Applications}\ }\textbf {\bibinfo {volume} {5}},\ \bibinfo
  {pages} {1} (\bibinfo {year} {2021}{\natexlab{a}})}\BibitemShut {NoStop}%
\bibitem [{\citenamefont {Cao}\ \emph {et~al.}(2021)\citenamefont {Cao},
  \citenamefont {Zhou}, \citenamefont {Wang}, \citenamefont {Ang},\ and\
  \citenamefont {Ang}}]{cao2021two}%
  \BibitemOpen
  \bibfield  {author} {\bibinfo {author} {\bibfnamefont {L.}~\bibnamefont
  {Cao}}, \bibinfo {author} {\bibfnamefont {G.}~\bibnamefont {Zhou}}, \bibinfo
  {author} {\bibfnamefont {Q.}~\bibnamefont {Wang}}, \bibinfo {author}
  {\bibfnamefont {L.}~\bibnamefont {Ang}}, \ and\ \bibinfo {author}
  {\bibfnamefont {Y.~S.}\ \bibnamefont {Ang}},\ }\href@noop {} {\bibfield
  {journal} {\bibinfo  {journal} {Applied Physics Letters}\ }\textbf {\bibinfo
  {volume} {118}},\ \bibinfo {pages} {013106} (\bibinfo {year}
  {2021})}\BibitemShut {NoStop}%
\bibitem [{\citenamefont {Yu}\ \emph {et~al.}(2021)\citenamefont {Yu},
  \citenamefont {Zhou}, \citenamefont {Wan},\ and\ \citenamefont
  {Li}}]{Yu2021}%
  \BibitemOpen
  \bibfield  {author} {\bibinfo {author} {\bibfnamefont {J.}~\bibnamefont
  {Yu}}, \bibinfo {author} {\bibfnamefont {J.}~\bibnamefont {Zhou}}, \bibinfo
  {author} {\bibfnamefont {X.}~\bibnamefont {Wan}}, \ and\ \bibinfo {author}
  {\bibfnamefont {Q.}~\bibnamefont {Li}},\ }\href {\doibase
  10.1088/1367-2630/abe8f7} {\bibfield  {journal} {\bibinfo  {journal} {New J.
  Phys}\ }\textbf {\bibinfo {volume} {23}},\ \bibinfo {pages} {33005} (\bibinfo
  {year} {2021})}\BibitemShut {NoStop}%
\bibitem [{\citenamefont {Wang}\ \emph
  {et~al.}(2021{\natexlab{b}})\citenamefont {Wang}, \citenamefont {Shi},
  \citenamefont {Liu}, \citenamefont {Zhang}, \citenamefont {Hong},
  \citenamefont {Li}, \citenamefont {Gao}, \citenamefont {Chen}, \citenamefont
  {Ren}, \citenamefont {Cheng} \emph {et~al.}}]{wang2021intercalated}%
  \BibitemOpen
  \bibfield  {author} {\bibinfo {author} {\bibfnamefont {L.}~\bibnamefont
  {Wang}}, \bibinfo {author} {\bibfnamefont {Y.}~\bibnamefont {Shi}}, \bibinfo
  {author} {\bibfnamefont {M.}~\bibnamefont {Liu}}, \bibinfo {author}
  {\bibfnamefont {A.}~\bibnamefont {Zhang}}, \bibinfo {author} {\bibfnamefont
  {Y.-L.}\ \bibnamefont {Hong}}, \bibinfo {author} {\bibfnamefont
  {R.}~\bibnamefont {Li}}, \bibinfo {author} {\bibfnamefont {Q.}~\bibnamefont
  {Gao}}, \bibinfo {author} {\bibfnamefont {M.}~\bibnamefont {Chen}}, \bibinfo
  {author} {\bibfnamefont {W.}~\bibnamefont {Ren}}, \bibinfo {author}
  {\bibfnamefont {H.-M.}\ \bibnamefont {Cheng}},  \emph {et~al.},\ }\href@noop
  {} {\bibfield  {journal} {\bibinfo  {journal} {Nature communications}\
  }\textbf {\bibinfo {volume} {12}},\ \bibinfo {pages} {1} (\bibinfo {year}
  {2021}{\natexlab{b}})}\BibitemShut {NoStop}%
\bibitem [{\citenamefont {Nandan}\ \emph {et~al.}(2021)\citenamefont {Nandan},
  \citenamefont {Ghosh}, \citenamefont {Agarwal}, \citenamefont {Bhowmick},\
  and\ \citenamefont {Chauhan}}]{nandan2021two}%
  \BibitemOpen
  \bibfield  {author} {\bibinfo {author} {\bibfnamefont {K.}~\bibnamefont
  {Nandan}}, \bibinfo {author} {\bibfnamefont {B.}~\bibnamefont {Ghosh}},
  \bibinfo {author} {\bibfnamefont {A.}~\bibnamefont {Agarwal}}, \bibinfo
  {author} {\bibfnamefont {S.}~\bibnamefont {Bhowmick}}, \ and\ \bibinfo
  {author} {\bibfnamefont {Y.~S.}\ \bibnamefont {Chauhan}},\ }\href@noop {}
  {\bibfield  {journal} {\bibinfo  {journal} {IEEE Transactions on Electron
  Devices}\ }\textbf {\bibinfo {volume} {69}},\ \bibinfo {pages} {406}
  (\bibinfo {year} {2021})}\BibitemShut {NoStop}%
\bibitem [{\citenamefont {Zeng}\ \emph {et~al.}(2012)\citenamefont {Zeng},
  \citenamefont {Dai}, \citenamefont {Yao}, \citenamefont {Xiao},\ and\
  \citenamefont {Cui}}]{zeng2012valley}%
  \BibitemOpen
  \bibfield  {author} {\bibinfo {author} {\bibfnamefont {H.}~\bibnamefont
  {Zeng}}, \bibinfo {author} {\bibfnamefont {J.}~\bibnamefont {Dai}}, \bibinfo
  {author} {\bibfnamefont {W.}~\bibnamefont {Yao}}, \bibinfo {author}
  {\bibfnamefont {D.}~\bibnamefont {Xiao}}, \ and\ \bibinfo {author}
  {\bibfnamefont {X.}~\bibnamefont {Cui}},\ }\href@noop {} {\bibfield
  {journal} {\bibinfo  {journal} {Nature {N}anotechnology}\ }\textbf {\bibinfo
  {volume} {7}},\ \bibinfo {pages} {490} (\bibinfo {year} {2012})}\BibitemShut
  {NoStop}%
\bibitem [{\citenamefont {Carvalho}\ \emph {et~al.}(2017)\citenamefont
  {Carvalho}, \citenamefont {Wang}, \citenamefont {Mignuzzi}, \citenamefont
  {Roy}, \citenamefont {Terrones}, \citenamefont {Fantini}, \citenamefont
  {Crespi}, \citenamefont {Malard},\ and\ \citenamefont
  {Pimenta}}]{carvalho2017intervalley}%
  \BibitemOpen
  \bibfield  {author} {\bibinfo {author} {\bibfnamefont {B.~R.}\ \bibnamefont
  {Carvalho}}, \bibinfo {author} {\bibfnamefont {Y.}~\bibnamefont {Wang}},
  \bibinfo {author} {\bibfnamefont {S.}~\bibnamefont {Mignuzzi}}, \bibinfo
  {author} {\bibfnamefont {D.}~\bibnamefont {Roy}}, \bibinfo {author}
  {\bibfnamefont {M.}~\bibnamefont {Terrones}}, \bibinfo {author}
  {\bibfnamefont {C.}~\bibnamefont {Fantini}}, \bibinfo {author} {\bibfnamefont
  {V.~H.}\ \bibnamefont {Crespi}}, \bibinfo {author} {\bibfnamefont {L.~M.}\
  \bibnamefont {Malard}}, \ and\ \bibinfo {author} {\bibfnamefont {M.~A.}\
  \bibnamefont {Pimenta}},\ }\href@noop {} {\bibfield  {journal} {\bibinfo
  {journal} {Nature {C}ommunications}\ }\textbf {\bibinfo {volume} {8}},\
  \bibinfo {pages} {1} (\bibinfo {year} {2017})}\BibitemShut {NoStop}%
\bibitem [{\citenamefont {Cao}\ \emph {et~al.}(2012)\citenamefont {Cao},
  \citenamefont {Wang}, \citenamefont {Han}, \citenamefont {Ye}, \citenamefont
  {Zhu}, \citenamefont {Shi}, \citenamefont {Niu}, \citenamefont {Tan},
  \citenamefont {Wang}, \citenamefont {Liu} \emph {et~al.}}]{cao2012valley}%
  \BibitemOpen
  \bibfield  {author} {\bibinfo {author} {\bibfnamefont {T.}~\bibnamefont
  {Cao}}, \bibinfo {author} {\bibfnamefont {G.}~\bibnamefont {Wang}}, \bibinfo
  {author} {\bibfnamefont {W.}~\bibnamefont {Han}}, \bibinfo {author}
  {\bibfnamefont {H.}~\bibnamefont {Ye}}, \bibinfo {author} {\bibfnamefont
  {C.}~\bibnamefont {Zhu}}, \bibinfo {author} {\bibfnamefont {J.}~\bibnamefont
  {Shi}}, \bibinfo {author} {\bibfnamefont {Q.}~\bibnamefont {Niu}}, \bibinfo
  {author} {\bibfnamefont {P.}~\bibnamefont {Tan}}, \bibinfo {author}
  {\bibfnamefont {E.}~\bibnamefont {Wang}}, \bibinfo {author} {\bibfnamefont
  {B.}~\bibnamefont {Liu}},  \emph {et~al.},\ }\href@noop {} {\bibfield
  {journal} {\bibinfo  {journal} {Nature {C}ommunications}\ }\textbf {\bibinfo
  {volume} {3}},\ \bibinfo {pages} {1} (\bibinfo {year} {2012})}\BibitemShut
  {NoStop}%
\bibitem [{\citenamefont {Wang}\ \emph {et~al.}(2012)\citenamefont {Wang},
  \citenamefont {Kalantar-Zadeh}, \citenamefont {Kis}, \citenamefont
  {Coleman},\ and\ \citenamefont {Strano}}]{wang2012electronics}%
  \BibitemOpen
  \bibfield  {author} {\bibinfo {author} {\bibfnamefont {Q.~H.}\ \bibnamefont
  {Wang}}, \bibinfo {author} {\bibfnamefont {K.}~\bibnamefont
  {Kalantar-Zadeh}}, \bibinfo {author} {\bibfnamefont {A.}~\bibnamefont {Kis}},
  \bibinfo {author} {\bibfnamefont {J.~N.}\ \bibnamefont {Coleman}}, \ and\
  \bibinfo {author} {\bibfnamefont {M.~S.}\ \bibnamefont {Strano}},\
  }\href@noop {} {\bibfield  {journal} {\bibinfo  {journal} {Nature
  nanotechnology}\ }\textbf {\bibinfo {volume} {7}},\ \bibinfo {pages} {699}
  (\bibinfo {year} {2012})}\BibitemShut {NoStop}%
\bibitem [{\citenamefont {He}\ \emph {et~al.}(2014)\citenamefont {He},
  \citenamefont {Kumar}, \citenamefont {Zhao}, \citenamefont {Wang},
  \citenamefont {Mak}, \citenamefont {Zhao},\ and\ \citenamefont
  {Shan}}]{he2014tightly}%
  \BibitemOpen
  \bibfield  {author} {\bibinfo {author} {\bibfnamefont {K.}~\bibnamefont
  {He}}, \bibinfo {author} {\bibfnamefont {N.}~\bibnamefont {Kumar}}, \bibinfo
  {author} {\bibfnamefont {L.}~\bibnamefont {Zhao}}, \bibinfo {author}
  {\bibfnamefont {Z.}~\bibnamefont {Wang}}, \bibinfo {author} {\bibfnamefont
  {K.~F.}\ \bibnamefont {Mak}}, \bibinfo {author} {\bibfnamefont
  {H.}~\bibnamefont {Zhao}}, \ and\ \bibinfo {author} {\bibfnamefont
  {J.}~\bibnamefont {Shan}},\ }\href@noop {} {\bibfield  {journal} {\bibinfo
  {journal} {Physical {R}eview {L}etters}\ }\textbf {\bibinfo {volume} {113}},\
  \bibinfo {pages} {026803} (\bibinfo {year} {2014})}\BibitemShut {NoStop}%
\bibitem [{\citenamefont {Wu}\ \emph {et~al.}(2013)\citenamefont {Wu},
  \citenamefont {Ross}, \citenamefont {Liu}, \citenamefont {Aivazian},
  \citenamefont {Jones}, \citenamefont {Fei}, \citenamefont {Zhu},
  \citenamefont {Xiao}, \citenamefont {Yao}, \citenamefont {Cobden} \emph
  {et~al.}}]{wu2013electrical}%
  \BibitemOpen
  \bibfield  {author} {\bibinfo {author} {\bibfnamefont {S.}~\bibnamefont
  {Wu}}, \bibinfo {author} {\bibfnamefont {J.~S.}\ \bibnamefont {Ross}},
  \bibinfo {author} {\bibfnamefont {G.-B.}\ \bibnamefont {Liu}}, \bibinfo
  {author} {\bibfnamefont {G.}~\bibnamefont {Aivazian}}, \bibinfo {author}
  {\bibfnamefont {A.}~\bibnamefont {Jones}}, \bibinfo {author} {\bibfnamefont
  {Z.}~\bibnamefont {Fei}}, \bibinfo {author} {\bibfnamefont {W.}~\bibnamefont
  {Zhu}}, \bibinfo {author} {\bibfnamefont {D.}~\bibnamefont {Xiao}}, \bibinfo
  {author} {\bibfnamefont {W.}~\bibnamefont {Yao}}, \bibinfo {author}
  {\bibfnamefont {D.}~\bibnamefont {Cobden}},  \emph {et~al.},\ }\href@noop {}
  {\bibfield  {journal} {\bibinfo  {journal} {Nature {P}hysics}\ }\textbf
  {\bibinfo {volume} {9}},\ \bibinfo {pages} {149} (\bibinfo {year}
  {2013})}\BibitemShut {NoStop}%
\bibitem [{\citenamefont {Jones}\ \emph {et~al.}(2013)\citenamefont {Jones},
  \citenamefont {Yu}, \citenamefont {Ghimire}, \citenamefont {Wu},
  \citenamefont {Aivazian}, \citenamefont {Ross}, \citenamefont {Zhao},
  \citenamefont {Yan}, \citenamefont {Mandrus}, \citenamefont {Xiao} \emph
  {et~al.}}]{jones2013optical}%
  \BibitemOpen
  \bibfield  {author} {\bibinfo {author} {\bibfnamefont {A.~M.}\ \bibnamefont
  {Jones}}, \bibinfo {author} {\bibfnamefont {H.}~\bibnamefont {Yu}}, \bibinfo
  {author} {\bibfnamefont {N.~J.}\ \bibnamefont {Ghimire}}, \bibinfo {author}
  {\bibfnamefont {S.}~\bibnamefont {Wu}}, \bibinfo {author} {\bibfnamefont
  {G.}~\bibnamefont {Aivazian}}, \bibinfo {author} {\bibfnamefont {J.~S.}\
  \bibnamefont {Ross}}, \bibinfo {author} {\bibfnamefont {B.}~\bibnamefont
  {Zhao}}, \bibinfo {author} {\bibfnamefont {J.}~\bibnamefont {Yan}}, \bibinfo
  {author} {\bibfnamefont {D.~G.}\ \bibnamefont {Mandrus}}, \bibinfo {author}
  {\bibfnamefont {D.}~\bibnamefont {Xiao}},  \emph {et~al.},\ }\href@noop {}
  {\bibfield  {journal} {\bibinfo  {journal} {Nature {N}anotechnology}\
  }\textbf {\bibinfo {volume} {8}},\ \bibinfo {pages} {634} (\bibinfo {year}
  {2013})}\BibitemShut {NoStop}%
\bibitem [{\citenamefont {Xu}\ \emph {et~al.}(2014)\citenamefont {Xu},
  \citenamefont {Yao}, \citenamefont {Xiao},\ and\ \citenamefont
  {Heinz}}]{xu2014spin}%
  \BibitemOpen
  \bibfield  {author} {\bibinfo {author} {\bibfnamefont {X.}~\bibnamefont
  {Xu}}, \bibinfo {author} {\bibfnamefont {W.}~\bibnamefont {Yao}}, \bibinfo
  {author} {\bibfnamefont {D.}~\bibnamefont {Xiao}}, \ and\ \bibinfo {author}
  {\bibfnamefont {T.~F.}\ \bibnamefont {Heinz}},\ }\href@noop {} {\bibfield
  {journal} {\bibinfo  {journal} {Nature Physics}\ }\textbf {\bibinfo {volume}
  {10}},\ \bibinfo {pages} {343} (\bibinfo {year} {2014})}\BibitemShut
  {NoStop}%
\bibitem [{\citenamefont {Chen}\ \emph
  {et~al.}(2021{\natexlab{b}})\citenamefont {Chen}, \citenamefont {Chen},
  \citenamefont {Deng}, \citenamefont {Watanabe}, \citenamefont {Taniguchi},
  \citenamefont {Huang},\ and\ \citenamefont {Xia}}]{chen2021probing}%
  \BibitemOpen
  \bibfield  {author} {\bibinfo {author} {\bibfnamefont {C.}~\bibnamefont
  {Chen}}, \bibinfo {author} {\bibfnamefont {X.}~\bibnamefont {Chen}}, \bibinfo
  {author} {\bibfnamefont {B.}~\bibnamefont {Deng}}, \bibinfo {author}
  {\bibfnamefont {K.}~\bibnamefont {Watanabe}}, \bibinfo {author}
  {\bibfnamefont {T.}~\bibnamefont {Taniguchi}}, \bibinfo {author}
  {\bibfnamefont {S.}~\bibnamefont {Huang}}, \ and\ \bibinfo {author}
  {\bibfnamefont {F.}~\bibnamefont {Xia}},\ }\href@noop {} {\bibfield
  {journal} {\bibinfo  {journal} {Physical {R}eview {B}}\ }\textbf {\bibinfo
  {volume} {103}},\ \bibinfo {pages} {035405} (\bibinfo {year}
  {2021}{\natexlab{b}})}\BibitemShut {NoStop}%
\bibitem [{\citenamefont {Drapcho}\ \emph {et~al.}(2017)\citenamefont
  {Drapcho}, \citenamefont {Kim}, \citenamefont {Hong}, \citenamefont {Jin},
  \citenamefont {Shi}, \citenamefont {Tongay}, \citenamefont {Wu},\ and\
  \citenamefont {Wang}}]{drapcho2017apparent}%
  \BibitemOpen
  \bibfield  {author} {\bibinfo {author} {\bibfnamefont {S.~G.}\ \bibnamefont
  {Drapcho}}, \bibinfo {author} {\bibfnamefont {J.}~\bibnamefont {Kim}},
  \bibinfo {author} {\bibfnamefont {X.}~\bibnamefont {Hong}}, \bibinfo {author}
  {\bibfnamefont {C.}~\bibnamefont {Jin}}, \bibinfo {author} {\bibfnamefont
  {S.}~\bibnamefont {Shi}}, \bibinfo {author} {\bibfnamefont {S.}~\bibnamefont
  {Tongay}}, \bibinfo {author} {\bibfnamefont {J.}~\bibnamefont {Wu}}, \ and\
  \bibinfo {author} {\bibfnamefont {F.}~\bibnamefont {Wang}},\ }\href@noop {}
  {\bibfield  {journal} {\bibinfo  {journal} {Physical {R}eview {B}}\ }\textbf
  {\bibinfo {volume} {95}},\ \bibinfo {pages} {165417} (\bibinfo {year}
  {2017})}\BibitemShut {NoStop}%
\bibitem [{\citenamefont {Tatsumi}\ and\ \citenamefont
  {Saito}(2018)}]{tatsumi2018interplay}%
  \BibitemOpen
  \bibfield  {author} {\bibinfo {author} {\bibfnamefont {Y.}~\bibnamefont
  {Tatsumi}}\ and\ \bibinfo {author} {\bibfnamefont {R.}~\bibnamefont
  {Saito}},\ }\href@noop {} {\bibfield  {journal} {\bibinfo  {journal}
  {Physical {R}eview {B}}\ }\textbf {\bibinfo {volume} {97}},\ \bibinfo {pages}
  {115407} (\bibinfo {year} {2018})}\BibitemShut {NoStop}%
\bibitem [{\citenamefont {Malard}\ \emph {et~al.}(2009)\citenamefont {Malard},
  \citenamefont {Pimenta}, \citenamefont {Dresselhaus},\ and\ \citenamefont
  {Dresselhaus}}]{malard2009raman}%
  \BibitemOpen
  \bibfield  {author} {\bibinfo {author} {\bibfnamefont {L.}~\bibnamefont
  {Malard}}, \bibinfo {author} {\bibfnamefont {M.~A.}\ \bibnamefont {Pimenta}},
  \bibinfo {author} {\bibfnamefont {G.}~\bibnamefont {Dresselhaus}}, \ and\
  \bibinfo {author} {\bibfnamefont {M.}~\bibnamefont {Dresselhaus}},\
  }\href@noop {} {\bibfield  {journal} {\bibinfo  {journal} {Physics reports}\
  }\textbf {\bibinfo {volume} {473}},\ \bibinfo {pages} {51} (\bibinfo {year}
  {2009})}\BibitemShut {NoStop}%
\bibitem [{\citenamefont {Zhu}\ \emph {et~al.}(2018)\citenamefont {Zhu},
  \citenamefont {Yi}, \citenamefont {Li}, \citenamefont {Xiao}, \citenamefont
  {Zhang}, \citenamefont {Yang}, \citenamefont {Kaindl}, \citenamefont {Li},
  \citenamefont {Wang},\ and\ \citenamefont {Zhang}}]{zhu2018observation}%
  \BibitemOpen
  \bibfield  {author} {\bibinfo {author} {\bibfnamefont {H.}~\bibnamefont
  {Zhu}}, \bibinfo {author} {\bibfnamefont {J.}~\bibnamefont {Yi}}, \bibinfo
  {author} {\bibfnamefont {M.-Y.}\ \bibnamefont {Li}}, \bibinfo {author}
  {\bibfnamefont {J.}~\bibnamefont {Xiao}}, \bibinfo {author} {\bibfnamefont
  {L.}~\bibnamefont {Zhang}}, \bibinfo {author} {\bibfnamefont {C.-W.}\
  \bibnamefont {Yang}}, \bibinfo {author} {\bibfnamefont {R.~A.}\ \bibnamefont
  {Kaindl}}, \bibinfo {author} {\bibfnamefont {L.-J.}\ \bibnamefont {Li}},
  \bibinfo {author} {\bibfnamefont {Y.}~\bibnamefont {Wang}}, \ and\ \bibinfo
  {author} {\bibfnamefont {X.}~\bibnamefont {Zhang}},\ }\href@noop {}
  {\bibfield  {journal} {\bibinfo  {journal} {Science}\ }\textbf {\bibinfo
  {volume} {359}},\ \bibinfo {pages} {579} (\bibinfo {year}
  {2018})}\BibitemShut {NoStop}%
\bibitem [{\citenamefont {Chen}\ \emph {et~al.}(2019)\citenamefont {Chen},
  \citenamefont {Lu}, \citenamefont {Dubey}, \citenamefont {Yao}, \citenamefont
  {Liu}, \citenamefont {Wang}, \citenamefont {Xiong}, \citenamefont {Zhang},\
  and\ \citenamefont {Srivastava}}]{chen2019entanglement}%
  \BibitemOpen
  \bibfield  {author} {\bibinfo {author} {\bibfnamefont {X.}~\bibnamefont
  {Chen}}, \bibinfo {author} {\bibfnamefont {X.}~\bibnamefont {Lu}}, \bibinfo
  {author} {\bibfnamefont {S.}~\bibnamefont {Dubey}}, \bibinfo {author}
  {\bibfnamefont {Q.}~\bibnamefont {Yao}}, \bibinfo {author} {\bibfnamefont
  {S.}~\bibnamefont {Liu}}, \bibinfo {author} {\bibfnamefont {X.}~\bibnamefont
  {Wang}}, \bibinfo {author} {\bibfnamefont {Q.}~\bibnamefont {Xiong}},
  \bibinfo {author} {\bibfnamefont {L.}~\bibnamefont {Zhang}}, \ and\ \bibinfo
  {author} {\bibfnamefont {A.}~\bibnamefont {Srivastava}},\ }\href@noop {}
  {\bibfield  {journal} {\bibinfo  {journal} {Nature Physics}\ }\textbf
  {\bibinfo {volume} {15}},\ \bibinfo {pages} {221} (\bibinfo {year}
  {2019})}\BibitemShut {NoStop}%
\bibitem [{\citenamefont {Li}\ \emph {et~al.}(2019{\natexlab{a}})\citenamefont
  {Li}, \citenamefont {Wang}, \citenamefont {Jin}, \citenamefont {Lu},
  \citenamefont {Lian}, \citenamefont {Meng}, \citenamefont {Blei},
  \citenamefont {Gao}, \citenamefont {Taniguchi}, \citenamefont {Watanabe}
  \emph {et~al.}}]{li2019momentum}%
  \BibitemOpen
  \bibfield  {author} {\bibinfo {author} {\bibfnamefont {Z.}~\bibnamefont
  {Li}}, \bibinfo {author} {\bibfnamefont {T.}~\bibnamefont {Wang}}, \bibinfo
  {author} {\bibfnamefont {C.}~\bibnamefont {Jin}}, \bibinfo {author}
  {\bibfnamefont {Z.}~\bibnamefont {Lu}}, \bibinfo {author} {\bibfnamefont
  {Z.}~\bibnamefont {Lian}}, \bibinfo {author} {\bibfnamefont {Y.}~\bibnamefont
  {Meng}}, \bibinfo {author} {\bibfnamefont {M.}~\bibnamefont {Blei}}, \bibinfo
  {author} {\bibfnamefont {M.}~\bibnamefont {Gao}}, \bibinfo {author}
  {\bibfnamefont {T.}~\bibnamefont {Taniguchi}}, \bibinfo {author}
  {\bibfnamefont {K.}~\bibnamefont {Watanabe}},  \emph {et~al.},\ }\href@noop
  {} {\bibfield  {journal} {\bibinfo  {journal} {{ACS} {N}ano}\ }\textbf
  {\bibinfo {volume} {13}},\ \bibinfo {pages} {14107} (\bibinfo {year}
  {2019}{\natexlab{a}})}\BibitemShut {NoStop}%
\bibitem [{\citenamefont {Li}\ \emph {et~al.}(2019{\natexlab{b}})\citenamefont
  {Li}, \citenamefont {Wang}, \citenamefont {Jin}, \citenamefont {Lu},
  \citenamefont {Lian}, \citenamefont {Meng}, \citenamefont {Blei},
  \citenamefont {Gao}, \citenamefont {Taniguchi}, \citenamefont {Watanabe}
  \emph {et~al.}}]{li2019emerging}%
  \BibitemOpen
  \bibfield  {author} {\bibinfo {author} {\bibfnamefont {Z.}~\bibnamefont
  {Li}}, \bibinfo {author} {\bibfnamefont {T.}~\bibnamefont {Wang}}, \bibinfo
  {author} {\bibfnamefont {C.}~\bibnamefont {Jin}}, \bibinfo {author}
  {\bibfnamefont {Z.}~\bibnamefont {Lu}}, \bibinfo {author} {\bibfnamefont
  {Z.}~\bibnamefont {Lian}}, \bibinfo {author} {\bibfnamefont {Y.}~\bibnamefont
  {Meng}}, \bibinfo {author} {\bibfnamefont {M.}~\bibnamefont {Blei}}, \bibinfo
  {author} {\bibfnamefont {S.}~\bibnamefont {Gao}}, \bibinfo {author}
  {\bibfnamefont {T.}~\bibnamefont {Taniguchi}}, \bibinfo {author}
  {\bibfnamefont {K.}~\bibnamefont {Watanabe}},  \emph {et~al.},\ }\href@noop
  {} {\bibfield  {journal} {\bibinfo  {journal} {Nature {C}ommunications}\
  }\textbf {\bibinfo {volume} {10}},\ \bibinfo {pages} {1} (\bibinfo {year}
  {2019}{\natexlab{b}})}\BibitemShut {NoStop}%
\bibitem [{\citenamefont {Streib}(2021)}]{simon2021difference}%
  \BibitemOpen
  \bibfield  {author} {\bibinfo {author} {\bibfnamefont {S.}~\bibnamefont
  {Streib}},\ }\href {\doibase 10.1103/PhysRevB.103.L100409} {\bibfield
  {journal} {\bibinfo  {journal} {Phys. {R}ev. {B}}\ }\textbf {\bibinfo
  {volume} {103}},\ \bibinfo {pages} {L100409} (\bibinfo {year}
  {2021})}\BibitemShut {NoStop}%
\bibitem [{\citenamefont {Zhang}\ and\ \citenamefont
  {Niu}(2015)}]{zhang2015chiral}%
  \BibitemOpen
  \bibfield  {author} {\bibinfo {author} {\bibfnamefont {L.}~\bibnamefont
  {Zhang}}\ and\ \bibinfo {author} {\bibfnamefont {Q.}~\bibnamefont {Niu}},\
  }\href@noop {} {\bibfield  {journal} {\bibinfo  {journal} {Physical {R}eview
  {L}etters}\ }\textbf {\bibinfo {volume} {115}},\ \bibinfo {pages} {115502}
  (\bibinfo {year} {2015})}\BibitemShut {NoStop}%
\bibitem [{\citenamefont {Chen}\ \emph
  {et~al.}(2021{\natexlab{c}})\citenamefont {Chen}, \citenamefont {Wu},
  \citenamefont {Zhu}, \citenamefont {Yang},\ and\ \citenamefont
  {Zhang}}]{chen2021propagating}%
  \BibitemOpen
  \bibfield  {author} {\bibinfo {author} {\bibfnamefont {H.}~\bibnamefont
  {Chen}}, \bibinfo {author} {\bibfnamefont {W.}~\bibnamefont {Wu}}, \bibinfo
  {author} {\bibfnamefont {J.}~\bibnamefont {Zhu}}, \bibinfo {author}
  {\bibfnamefont {S.~A.}\ \bibnamefont {Yang}}, \ and\ \bibinfo {author}
  {\bibfnamefont {L.}~\bibnamefont {Zhang}},\ }\href@noop {} {\bibfield
  {journal} {\bibinfo  {journal} {Nano {L}etters}\ }\textbf {\bibinfo {volume}
  {21}},\ \bibinfo {pages} {3060} (\bibinfo {year}
  {2021}{\natexlab{c}})}\BibitemShut {NoStop}%
\bibitem [{\citenamefont {Zhang}\ and\ \citenamefont
  {Murakami}(2022)}]{zhang2021chiral}%
  \BibitemOpen
  \bibfield  {author} {\bibinfo {author} {\bibfnamefont {T.}~\bibnamefont
  {Zhang}}\ and\ \bibinfo {author} {\bibfnamefont {S.}~\bibnamefont
  {Murakami}},\ }\href {\doibase 10.1103/PhysRevResearch.4.L012024} {\bibfield
  {journal} {\bibinfo  {journal} {Phys. {R}ev. {R}esearch}\ }\textbf {\bibinfo
  {volume} {4}},\ \bibinfo {pages} {L012024} (\bibinfo {year}
  {2022})}\BibitemShut {NoStop}%
\bibitem [{\citenamefont {Splendiani}\ \emph {et~al.}(2010)\citenamefont
  {Splendiani}, \citenamefont {Sun}, \citenamefont {Zhang}, \citenamefont {Li},
  \citenamefont {Kim}, \citenamefont {Chim}, \citenamefont {Galli},\ and\
  \citenamefont {Wang}}]{splendiani2010emerging}%
  \BibitemOpen
  \bibfield  {author} {\bibinfo {author} {\bibfnamefont {A.}~\bibnamefont
  {Splendiani}}, \bibinfo {author} {\bibfnamefont {L.}~\bibnamefont {Sun}},
  \bibinfo {author} {\bibfnamefont {Y.}~\bibnamefont {Zhang}}, \bibinfo
  {author} {\bibfnamefont {T.}~\bibnamefont {Li}}, \bibinfo {author}
  {\bibfnamefont {J.}~\bibnamefont {Kim}}, \bibinfo {author} {\bibfnamefont
  {C.-Y.}\ \bibnamefont {Chim}}, \bibinfo {author} {\bibfnamefont
  {G.}~\bibnamefont {Galli}}, \ and\ \bibinfo {author} {\bibfnamefont
  {F.}~\bibnamefont {Wang}},\ }\href@noop {} {\bibfield  {journal} {\bibinfo
  {journal} {Nano letters}\ }\textbf {\bibinfo {volume} {10}},\ \bibinfo
  {pages} {1271} (\bibinfo {year} {2010})}\BibitemShut {NoStop}%
\bibitem [{\citenamefont {Xiao}\ \emph {et~al.}(2012)\citenamefont {Xiao},
  \citenamefont {Liu}, \citenamefont {Feng}, \citenamefont {Xu},\ and\
  \citenamefont {Yao}}]{xiao2012coupled}%
  \BibitemOpen
  \bibfield  {author} {\bibinfo {author} {\bibfnamefont {D.}~\bibnamefont
  {Xiao}}, \bibinfo {author} {\bibfnamefont {G.-B.}\ \bibnamefont {Liu}},
  \bibinfo {author} {\bibfnamefont {W.}~\bibnamefont {Feng}}, \bibinfo {author}
  {\bibfnamefont {X.}~\bibnamefont {Xu}}, \ and\ \bibinfo {author}
  {\bibfnamefont {W.}~\bibnamefont {Yao}},\ }\href@noop {} {\bibfield
  {journal} {\bibinfo  {journal} {Physical {R}eview {L}etters}\ }\textbf
  {\bibinfo {volume} {108}},\ \bibinfo {pages} {196802} (\bibinfo {year}
  {2012})}\BibitemShut {NoStop}%
\bibitem [{\citenamefont {Yao}\ \emph {et~al.}(2008)\citenamefont {Yao},
  \citenamefont {Xiao},\ and\ \citenamefont {Niu}}]{yao2008valley}%
  \BibitemOpen
  \bibfield  {author} {\bibinfo {author} {\bibfnamefont {W.}~\bibnamefont
  {Yao}}, \bibinfo {author} {\bibfnamefont {D.}~\bibnamefont {Xiao}}, \ and\
  \bibinfo {author} {\bibfnamefont {Q.}~\bibnamefont {Niu}},\ }\href@noop {}
  {\bibfield  {journal} {\bibinfo  {journal} {Physical {R}eview {B}}\ }\textbf
  {\bibinfo {volume} {77}},\ \bibinfo {pages} {235406} (\bibinfo {year}
  {2008})}\BibitemShut {NoStop}%
\bibitem [{\citenamefont {Kresse}\ and\ \citenamefont
  {Furthm{\"u}ller}(1996)}]{CAL_VASP}%
  \BibitemOpen
  \bibfield  {author} {\bibinfo {author} {\bibfnamefont {G.}~\bibnamefont
  {Kresse}}\ and\ \bibinfo {author} {\bibfnamefont {J.}~\bibnamefont
  {Furthm{\"u}ller}},\ }\href@noop {} {\bibfield  {journal} {\bibinfo
  {journal} {Physical {R}eview {B}}\ }\textbf {\bibinfo {volume} {54}},\
  \bibinfo {pages} {11169} (\bibinfo {year} {1996})}\BibitemShut {NoStop}%
\bibitem [{\citenamefont {Perdew}\ \emph {et~al.}(1996)\citenamefont {Perdew},
  \citenamefont {Burke},\ and\ \citenamefont {Ernzerhof}}]{DFT_GGA}%
  \BibitemOpen
  \bibfield  {author} {\bibinfo {author} {\bibfnamefont {J.~P.}\ \bibnamefont
  {Perdew}}, \bibinfo {author} {\bibfnamefont {K.}~\bibnamefont {Burke}}, \
  and\ \bibinfo {author} {\bibfnamefont {M.}~\bibnamefont {Ernzerhof}},\
  }\href@noop {} {\bibfield  {journal} {\bibinfo  {journal} {Physical {R}eview
  {L}etters}\ }\textbf {\bibinfo {volume} {77}},\ \bibinfo {pages} {3865}
  (\bibinfo {year} {1996})}\BibitemShut {NoStop}%
\bibitem [{\citenamefont {Grimme}\ \emph {et~al.}(2010)\citenamefont {Grimme},
  \citenamefont {Antony}, \citenamefont {Ehrlich},\ and\ \citenamefont
  {Krieg}}]{grimme2010consistent}%
  \BibitemOpen
  \bibfield  {author} {\bibinfo {author} {\bibfnamefont {S.}~\bibnamefont
  {Grimme}}, \bibinfo {author} {\bibfnamefont {J.}~\bibnamefont {Antony}},
  \bibinfo {author} {\bibfnamefont {S.}~\bibnamefont {Ehrlich}}, \ and\
  \bibinfo {author} {\bibfnamefont {H.}~\bibnamefont {Krieg}},\ }\href@noop {}
  {\bibfield  {journal} {\bibinfo  {journal} {The Journal of chemical physics}\
  }\textbf {\bibinfo {volume} {132}},\ \bibinfo {pages} {154104} (\bibinfo
  {year} {2010})}\BibitemShut {NoStop}%
\bibitem [{\citenamefont {Grimme}\ \emph {et~al.}(2011)\citenamefont {Grimme},
  \citenamefont {Ehrlich},\ and\ \citenamefont {Goerigk}}]{grimme2011effect}%
  \BibitemOpen
  \bibfield  {author} {\bibinfo {author} {\bibfnamefont {S.}~\bibnamefont
  {Grimme}}, \bibinfo {author} {\bibfnamefont {S.}~\bibnamefont {Ehrlich}}, \
  and\ \bibinfo {author} {\bibfnamefont {L.}~\bibnamefont {Goerigk}},\
  }\href@noop {} {\bibfield  {journal} {\bibinfo  {journal} {Journal of
  computational chemistry}\ }\textbf {\bibinfo {volume} {32}},\ \bibinfo
  {pages} {1456} (\bibinfo {year} {2011})}\BibitemShut {NoStop}%
\end{thebibliography}%

%dummy comment inserted by tex2lyx to ensure that this paragraph is not empty%dummy comment inserted by tex2lyx to ensure that this paragraph is not empty

 \newpage{}
\end{document}